\documentclass[journal]{IEEEtran}

\IEEEoverridecommandlockouts                              % This command is only
\usepackage{amssymb}
\usepackage{amsmath}
\usepackage{bm}
\usepackage{mathrsfs}
\usepackage{comment} 
\usepackage{algorithmic,algorithm}
\usepackage{shuffle}
\usepackage{graphicx} % for pdf, bitmapped graphics files
\usepackage{caption}
\usepackage{lipsum}
\usepackage{color}
\usepackage{enumerate}
\usepackage{amsfonts}
\usepackage{subfigure} 
\usepackage{array}
\usepackage{enumerate}
\usepackage{setspace}
\usepackage{multirow}
\DeclareSymbolFont{symbolsC}{U}{pxsyc}{m}{n}
\DeclareMathSymbol{\coloneqq}{\mathrel}{symbolsC}{"42}

\begin{document}

\title{\LARGE \bf
Synthesis of Maximally Permissive Covert  Attackers Against Unknown Supervisors by Using Observations} %Partially Observed  via Projections of Holons
\author{Ruochen Tai, Liyong Lin, Yuting Zhu and Rong Su
%\thanks{
%This work was partially supported by the National Natural Science Foundation of China under Grant
%Nos. 61374068, 61472295, and 61672400, the Recruitment Program of Global Experts, and the Science
%and Technology Development Fund, MSAR, under Grant Nos. 078/2015/A3 and 106/20156/A3
%(Corresponding author: Z. Li).

%D.~Wang is with the School of Electro-Mechanical Engineering, Xidian University, Xi'an 710071, China (e-mail: wdeguang1991@163.com).
%
%L.~Lin is with the Systems Control Group, Department of Electrical
%and Computer Engineering, University of Toronto, Toronto, ON M5S 3G4
%Canada (e-mail: liyong.lin@utoronto.ca).
%
%Z. Li is with the Institute of Systems Engineering, Macau University of Science and Technology, Taipa,
%Macau and also with the Key Laboratory of Electronic Equipment Structure Design, Ministry of
%Education, School of Electro-Mechanical Engineering, Xidian University, Xi'an 710071, China (e-mail: zhwli@xidian.edu.cn).
%
%W.~M.~Wonham is with the Systems Control Group, Department of Electrical
%and Computer Engineering, University of Toronto, Toronto, ON M5S 3G4,
%Canada (e-mail: wonham@control.utoronto.ca).}
%}
\thanks{The research of the project was supported by Ministry of Education, Singapore, under grant AcRF TIER 1-2018-T1-001-245 (RG 91/18).

The authors are affliated with Nanyang Technological University, Singapore. (Email: ruochen001@e.ntu.edu.sg; liyong.lin@ntu.edu.sg; yuting002@e.ntu.edu.sg; rsu@ntu.edu.sg). Ruochen Tai and Liyong Lin contribute equally to this work.
(\emph{Corresponding author: Liyong Lin})}
}
\maketitle

%\begin{comment}
\begin{abstract}
In this paper, we consider the problem of synthesis of maximally permissive covert  damage-reachable  attackers in the setup where the model of the supervisor is unknown to the adversary but the adversary has recorded a (prefix-closed) finite set of observations of the runs of the closed-loop system. The synthesized attacker needs to ensure both the damage-reachability and the covertness against all the supervisors which are consistent with the given set of observations. There is a gap between the de facto maximal permissiveness, assuming the model of the supervisor is known, and the maximal permissiveness that can be attained with a limited knowledge of the model of the supervisor, from the adversary's point of view. We consider the setup where the attacker can exercise sensor replacement/deletion attacks and actuator enablement/disablement attacks. The solution methodology proposed in this work is to reduce the synthesis of maximally permissive covert  damage-reachable attackers, given the model of the plant and the finite set of observations, to the synthesis of maximally permissive safe supervisors for certain transformed plant, which shows the decidability of the observation-assisted covert attacker synthesis problem.  The effectiveness of our approach is illustrated on a water tank example adapted from the literature. 
\end{abstract}


%{\color{red} safety and liveness, mu caluclus, buchi automaton}

%{\color{red} title may need change}

\section{Introduction}
\label{sec:intro}

The security of cyber-physical system, modelled in the abstraction level of events~\cite{WMW10}, has  attracted much research interest from the discrete-event system  community, with most of the existing works devoted to attack detection and security verification \cite{CarvalhoEnablementAttacks}-\cite{WP}, synthesis of covert attackers \cite{Goes2017}-\cite{Su2018}, and synthesis of resilient supervisors \cite{Su2018}-\cite{LS20BJ}. %In this paper, we shall focus on the synthesis of covert attackers, but without imposing the assumption that the model of the supervisor is available to the adversary, in contrast to the setup of \cite{Goes2017}-\cite{Su2018}.

%in a more practical setup than those of \cite{Goes2017}-\cite{Su2018}. In~\cite{Goes2017}-\cite{Su2018}, extending and improving our recent work \cite{LTZS20} in which the supervisor model is not assumed to be given a prior.

The problem of covert sensor attacker synthesis has been studied extensively \cite{Goes2017}-\cite{Mohajerani20}, \cite{Su2018}. In \cite{Su2018}, it is shown that,
under a normality assumption on the sensor attackers, the supremal covert sensor attacker  exists and can be effectively synthesized. In \cite{Goes2017, Goes2020}, a game-theoretic approach is presented to synthesize covert sensor attackers, without imposing the normality assumption. Recently, based on the game arena of \cite{Goes2017} and \cite{Goes2020}, \cite{Mohajerani20} develops an abstraction based synthesis approach to improve the synthesis efficiency. The problem of covert actuator attacker synthesis has been addressed in \cite{Lin2018} and \cite{LZS19}, by  employing a reduction to the (partial-observation) supervisor synthesis problem \cite{LZS19}. With the reduction based approach,  the more general problem of covert actuator and sensor attacker synthesis has also been addressed \cite{LS20}-\cite{Kh19}. The synthesis approaches developed in \cite{Goes2017}-\cite{Su2018} allow maximally permissive covert attackers to be synthesized from the model of the plant and the model of the supervisor. However, it can be restrictive to assume the model of the supervisor to be known to the adversary, which is unlikely unless the adversary is an insider.%, which may not hold in practice and render the synthesis procedures developed in~\cite{Su2018, Goes2017, Goes2020, LZS19, Lin2018, LS20, Kh19, LS20J, Mohajerani20} ineffective.  

Recently, we have considered a more practical  setup where the model of the supervisor is not available to the adversary \cite{LTZS20}.
To compensate the lack of knowledge on the model of the supervisor, it is assumed in \cite{LTZS20} that the adversary has recorded a (prefix-closed)  finite set of observations of the runs of the closed-loop system. In this more challenging setup, a covert attacker needs to be synthesized based solely on the model of the plant and the given  set of observations. And the synthesized attacker needs to ensure\footnote{From the adversary's point of view, any supervisor that is consistent with the given set of observations may have been deployed.} the damage-reachability and the covertness against all the supervisors that are consistent with the given set of observations. The difficulty of this synthesis problem lies in the fact that there can be in general an infinite number of supervisors which are consistent with the observations, rendering the synthesis approaches developed in the existing works ineffective. In \cite{LTZS20}, we have proposed a technique to compute covert  damage-reachable attackers by formulating it as an instance of the supervisor synthesis problem on certain surrogate plant model, which is constructed without using the model of the supervisor. Due to the  over-approximation  in the surrogate plant model, the synthesized attacker in \cite{LTZS20} cannot ensure the maximal permissiveness in general. It is worth noting that there is a gap between the de facto maximal permissiveness, assuming the model of the supervisor is known,  and the maximal permissiveness (from the adversary's point of view) that could be attained with a  limited knowledge of the model of the supervisor. It is the maximal permissiveness from the adversary's point of view that is of interest in this work.
%Due to the 
%The synthesis approach developed in the existing works is quite powerful, in the sense that maximally permissive covert attackers can be synthesized from the model of the plant and the model of the supervisor.   While it is  natural to assume the model of the plant to be known a prior, it seems a bit restrictive  to also assume the model of the supervisor to be  known, which limits the usefulness of the existing covert attacker synthesis procedures in practice. 
 %Indeed, in this setup, the synthesized attacker needs to ensure damage-infliction and covertness against all the supervisors that are consistent with the finite set of observations, which limits the maximal permissiveness that can be attained under such a model uncertainty. 
%However,  with a guarantee of damage-infliction and maximal permissiveness with respect to the observations. 
%It is of interest to understand 

In this paper, as in \cite{LTZS20}, we also assume the model of the supervisor is not available to the adversary and the adversary can use the observations to assist the synthesis of covert attackers. We consider attackers whose attack mechanisms are restricted to sensor replacement/deletion attacks and actuator enablement/disablement attacks in this work. The main contributions of this work are listed as follows.
\begin{itemize}
\setlength{\itemsep}{3pt}
\setlength{\parsep}{0pt}
\setlength{\parskip}{0pt}
    \item We provide a sound and complete procedure for the synthesis of covert  damage-reachable attackers, given the model of the plant and the finite set of observations. The solution methodology is to reduce it to the problem of partial-observation supervisor synthesis for certain transformed plant, which shows the decidability of the observation-assisted covert (damage-reachable) attacker synthesis problem. We allow sensor replacement/deletion attacks\footnote{It is also possible to deal with sensor insertion attacks by using our approach, which requires some modifications in our constructions. For simplicity, we will not address sensor insertion attacks in this work.} and actuator enablement/disablement attacks. In comparison, there are two limitations regarding the approach proposed in \cite{LTZS20}: 1) it only provides a sound, but generally incomplete, heuristic algorithm for the synthesis of covert damage-reachable attackers due to the use of  over-approximation in the surrogate plant, and 2) it cannot deal with actuator enablement attacks. 
    %Our approach is applicable for the setup where the supervisors and the attackers possess different observation capabilities on the plant, in contrast to \cite{LTZS20} where the supervisor and the attacker have the same observation capability on sensor events. 
    % (this seems not a serious limitation of [24]) In our solution methodology, there is also no need for the verification of the damage-reachability of the synthesized attackers, compared with \cite{LTZS20}. 
    \item The approach proposed in this work can synthesize maximally permissive covert damage-reachable attackers, among those attackers which can ensure the damage-reachability and the covertness against all the supervisors which are consistent with the set of observations. 
    We provide a formal proof of the maximal permissiveness and the correctness of the synthesized attackers, by reasoning on the model of the attacked closed-loop system, adapted from \cite{LS20}, \cite{LS20J}, \cite{LS20BJ}. In comparison, maximal permissiveness is not guaranteed in \cite{LTZS20}, due to the use of over-approximation in the surrogate plant. 
\end{itemize}

In practice, one may observe the closed-loop system for a sufficiently long time, i.e., obtain a sufficient number of observations of the runs of the closed-loop system, and hope to learn an exact observable model of the  closed-loop system, that is, the natural projection of the closed-loop system, also known as the monitor~\cite{LS20J}. However, this approach has two problems. First of all, it is not efficient, indeed infeasible, to learn the observable model of the  closed-loop system, as in theory an infinite number of runs needs to be observed. We can never guarantee the correctness of the learnt model for any finite set of observations, without an oracle for confirming the correctness of the learnt model. Secondly, even if we obtain an exact observable model of the closed-loop system, the model in general has insufficient information for us to  extract a model of the supervisor and use, for example, the technique developed in \cite{LZS19,LS20} for synthesizing covert attackers. A much more viable and efficient approach is to observe the closed-loop system for just long enough, by observing as few runs of the closed-loop system as possible, to extract just enough information to carry out the synthesis of an non-empty covert attacker. If a given set of observations is verified to be
sufficient for us to synthesize a non-empty covert attacker, then we know that more observations will only allow more permissive covert attacker to be synthesized. 
The solution proposed in this work can determine if any given set of observations contains enough information for the synthesis of a non-empty covert attacker and can directly synthesize a covert attacker from the set of observations whenever it is possible. %{\color{red} ?The same synthesis procedure also allows us to determine if a non-empty covert  attacker can be synthesized in the limit, even when all the possible observations can be collected. This can determine in each case if the use of observations for the synthesis of covert damage-reachable attacks is fundamentally infeasible, when the model of the supervisor is unknown.if it is possilble, can we always have a finite feasible set?}

This paper is organized as follows. In Section \ref{sec:Preliminaries}, we recall the preliminaries which are needed for understanding this paper. In Section \ref{sec:Component models under sensor-actuator attack}, we then introduce the system setup and present the model constructions. The proposed synthesis solution as well as the correctness proof are presented in Section \ref{sec:Synthesis of Maximally Permissive Covert Attackers Against Unknown Supervisors}. Finally, in Section \ref{sec:Conclusions}, the conclusions are drawn. A running example is given throughout the paper. 

%%%%%%%%%%%%%%%%%%%%%%%%%%%%%%%%%%%%%%%%%%%%%%%%%%%%%%%%%%%%%%%%%%%%%%%%%%%%%%%%%%%%

\section{Preliminaries}
\label{sec:Preliminaries}
In this section, we introduce some basic notations and terminologies that will be used in this work, mostly following~\cite{WMW10, CL99, HU79}.  %{\color{red} to edit} %and (quantified) Boolean formulas~\cite{BHM09}. 
 
%For any set $A$, we write $|A|$ to denote its cardinality. 
 
%For any two sets $A$ and $B$, we use $A \times B$ to denote their Cartesian product and use $A-B$ to denote their difference. For any relation $R \subseteq A \times B$ and any $a \in A$, we define $R[a]:=\{b \in B \mid (a, b) \in R\}$.

Given a finite alphabet $\Sigma$, let $\Sigma^{*}$ be the free monoid over $\Sigma$ with the empty string $\varepsilon$ being the unit element and the string concatenation being the monoid operation. For a string $s$, $|s|$ is defined to be the length of $s$. Given two strings $s, t \in \Sigma^{*}$, we say $s$ is a prefix substring of $t$, written as $s \leq t$, if there exists $u \in \Sigma^{*}$ such that $su = t$, where $su$ denotes the concatenation of $s$ and $u$. A language $L \subseteq \Sigma^{*}$ is a set of strings. The prefix closure of $L$ is defined as $\overline{L} = \{u \in \Sigma^{*} \mid (\exists v \in L) \, u\leq v\}$. 
%If $L = \overline{L}$, then $L$ is \emph{prefix-closed}. The concatenation of two languages $L_{a}, L_{b} \subseteq \Sigma^{*}$ is defined as $L_{a}L_{b} = \{s_{a}s_{b} \in \Sigma^{*}|s_{a} \in L_{a} \wedge s_{b} \in L_{b}\}$.
The event set $\Sigma$ is partitioned into $\Sigma = \Sigma_{c} \dot{\cup} \Sigma_{uc} = \Sigma_{o} \dot{\cup} \Sigma_{uo}$, where $\Sigma_{c}$ (respectively, $\Sigma_{o}$) and $\Sigma_{uc}$ (respectively, $\Sigma_{uo}$) are defined as the sets of controllable (respectively, observable) and uncontrollable (respectively, unobservable) events, respectively.  As usual, $P_{o}: \Sigma^{*} \rightarrow \Sigma_{o}^{*}$ is the natural projection defined such that
\begin{enumerate}[(1)]
\setlength{\itemsep}{3pt}
\setlength{\parsep}{0pt}
\setlength{\parskip}{0pt}
\item $P_{o}(\varepsilon) = \varepsilon$,
\item $(\forall \sigma \in \Sigma) \, P_{o}(\sigma)=
\left\{
\begin{array}{rcl}
\sigma       &      & {\sigma \in \Sigma_{o},}\\
\varepsilon  &      & {\rm otherwise,}
\end{array} \right.$
\item $(\forall s \in \Sigma^*, \sigma \in \Sigma) \, P_{o}(s\sigma) = P_{o}(s)P_{o}(\sigma)$.
\end{enumerate}
We sometimes also write $P_o$ as $P_{\Sigma_o}$, to explicitly illustrate the co-domain $\Sigma_o^*$.

A finite state automaton $G$ over $\Sigma$ is given by a 5-tuple $(Q, \Sigma, \xi, q_{0}, Q_{m})$, where $Q$ is the state set, $\xi: Q \times \Sigma \rightarrow Q$ is the (partial) transition function, $q_{0} \in Q$ is the initial state, and $Q_{m}$ is the set of marker states. 
We write $\xi(q, \sigma)!$ to mean that $\xi(q, \sigma)$ is defined and also view $\xi \subseteq Q \times \Sigma \times Q$ as a relation. We define $En_{G}(q) = \{\sigma \in \Sigma|\xi(q, \sigma)!\}$.
$\xi$ is also extended to the (partial) transition function $\xi: Q \times \Sigma^{*} \rightarrow Q$ and the transition function $\xi: 2^{Q} \times \Sigma \rightarrow 2^{Q}$ \cite{WMW10}, where the later is defined as follows: for any $Q' \subseteq Q$ and any $\sigma \in \Sigma$, $\xi(Q', \sigma) = \{q' \in Q|(\exists q \in Q')q' = \xi(q, \sigma)\}$. 
Let $L(G)$ and $L_{m}(G)$ denote the closed-behavior and the marked behavior, respectively. $G$ is said to be marker-reachable if some marker state of $G$ is reachable~\cite{WMW10}. $G$ is marker-reachable iff $L_m(G) \neq \varnothing$. When $Q_{m} = Q$, we shall also write $G = (Q, \Sigma, \xi, q_{0})$ for simplicity. 
The ``unobservable reach'' of the state $q \in Q$ under the subset of events $\Sigma' \subseteq \Sigma$ is given by $UR_{G, \Sigma - \Sigma'}(q) := \{q' \in Q|[\exists s \in (\Sigma - \Sigma')^{*}] \, q' = \xi(q,s)\}$.
We shall abuse the notation and define $P_{\Sigma'}(G)$ to be the finite state automaton $(2^{Q} - \{\varnothing\}, \Sigma, \delta, UR_{G, \Sigma - \Sigma'}(q_{0}))$ over $\Sigma$, where $UR_{G, \Sigma - \Sigma'}(q_{0}) \in 2^Q-\{\varnothing\}$ is the initial state, and the (partial) transition function $\delta: (2^{Q} - \{\varnothing\}) \times \Sigma \rightarrow (2^{Q} - \{\varnothing\})$ is defined as follows:
\begin{enumerate}[(1)]
    \setlength{\itemsep}{3pt}
    \setlength{\parsep}{0pt}
    \setlength{\parskip}{0pt}
    \item For any $\varnothing \neq Q' \subseteq Q$ and any $\sigma \in \Sigma'$, if $\xi(Q', \sigma) \neq \varnothing$, then $\delta(Q', \sigma) = UR_{G, \Sigma - \Sigma'}(\xi(Q', \sigma))$, where
    \[
    UR_{G, \Sigma - \Sigma'}(Q'') = \bigcup\limits_{q \in Q''}UR_{G, \Sigma - \Sigma'}(q)
    \]
    for any $\varnothing \neq Q'' \subseteq Q$.%if $\xi(Q', \sigma) = \varnothing$, then $\neg \delta(Q', \sigma)!$.
    \item For any $\varnothing \neq Q' \subseteq Q$ and any $\sigma \in \Sigma - \Sigma'$, $\delta(Q', \sigma) = Q'$.
\end{enumerate}
It is noteworthy that $P_{\Sigma'}(G)$ is over $\Sigma$, instead of $\Sigma'$ and here we exclude the state $\varnothing$ in its state space.
%\vspace{-0.3cm}

%A finite state automaton $G = (Q, \Sigma, \xi, q_{0}, Q_{m})$ is said to be non-blocking if every reachable state in $G$ can reach some marked state in $Q_{m}$ \cite{WMW10}. 
As usual, for any two finite state automata $G_{1} = (Q_{1}, \Sigma_{1}, \xi_{1}, q_{1,0}, Q_{1,m})$ and $G_{2} = (Q_{2}, \Sigma_{2}, \xi_{2}, q_{2,0}, Q_{2,m})$, where $En_{G_{1}}(q) = \{\sigma \in \Sigma_1|\xi_{1}(q, \sigma)!\}$ and $En_{G_{2}}(q) = \{\sigma \in \Sigma_2|\xi_{2}(q, \sigma)!\}$, their synchronous product \cite{CL99} is denoted as $G_{1}||G_{2} := (Q_{1} \times Q_{2}, \Sigma_{1} \cup \Sigma_{2}, \zeta, (q_{1,0}, q_{2,0}), Q_{1,m} \times Q_{2,m})$, where the (partial) transition function $\zeta$ is defined as follows, for any $(q_{1}, q_{2}) \in Q_{1} \times Q_{2}$ and $\sigma \in \Sigma = \Sigma_1 \cup \Sigma_2$:
\[
\begin{aligned}
& \zeta((q_{1}, q_{2}), \sigma) := \\ & \left\{
\begin{array}{lcl}
(\xi_{1}(q_{1}, \sigma), \xi_{2}(q_{2}, \sigma))  &      & {\rm if} \, {\sigma \in En_{G_{1}}(q_{1}) \cap En_{G_{2}}(q_{2}),}\\
(\xi_{1}(q_{1}, \sigma), q_{2})       &      & {\rm if} \, {\sigma \in En_{G_{1}}(q_{1}) \backslash \Sigma_{2},}\\
(q_{1}, \xi_{2}(q_{2}, \sigma))       &      & {\rm if} \, {\sigma \in En_{G_{2}}(q_{2}) \backslash \Sigma_{1},}\\
{\rm not \, defined}  &      & {\rm otherwise.}
\end{array} \right.
\end{aligned}
\]
For convenience, for any two finite state automata $G_{1}$ and $G_{2}$, we write $G_1=G_2$ iff $L(G_{1}) = L(G_{2})$ and $L_{m}(G_{1}) = L_{m}(G_{2})$. We also write $G_1 \sqsubseteq G_2$ iff $L(G_{1}) \subseteq L(G_{2})$ and $L_{m}(G_{1}) \subseteq  L_{m}(G_{2})$. It then follows that $G_1=G_2$ iff $G_1 \sqsubseteq G_2$ and $G_2 \sqsubseteq G_1$.

\textbf{Notation.} Let $\Gamma = \{\gamma \subseteq \Sigma|\Sigma_{uc} \subseteq \gamma\}$ denote the set of all the possible control commands. In this work, it is assumed that when no control command is received by plant $G$, then only uncontrollable events could be executed. 
For a set $\Sigma$, we use $\Sigma^{\#}$ to denote a copy of $\Sigma$ with superscript ``$\#$'' attached to each element in $\Sigma$. Intuitively speaking, ``$\#$'' denotes the message tampering due to the sensor attacks; the specific meanings of the relabelled events will be introduced later in Section \ref{sec:Component models under sensor-actuator attack}.

%%%%%%%%%%%%%%%%%%%%%%%%%%%%%%%%%%%%%%%%%%%%%%%%%%%%%%%%%%%%%%%%%%%%%%%%%%%%%%%%%%%%

\section{Component models under sensor-actuator attack}
\label{sec:Component models under sensor-actuator attack}

\begin{figure}[htbp]
\begin{center}
\includegraphics[height=4.7cm]{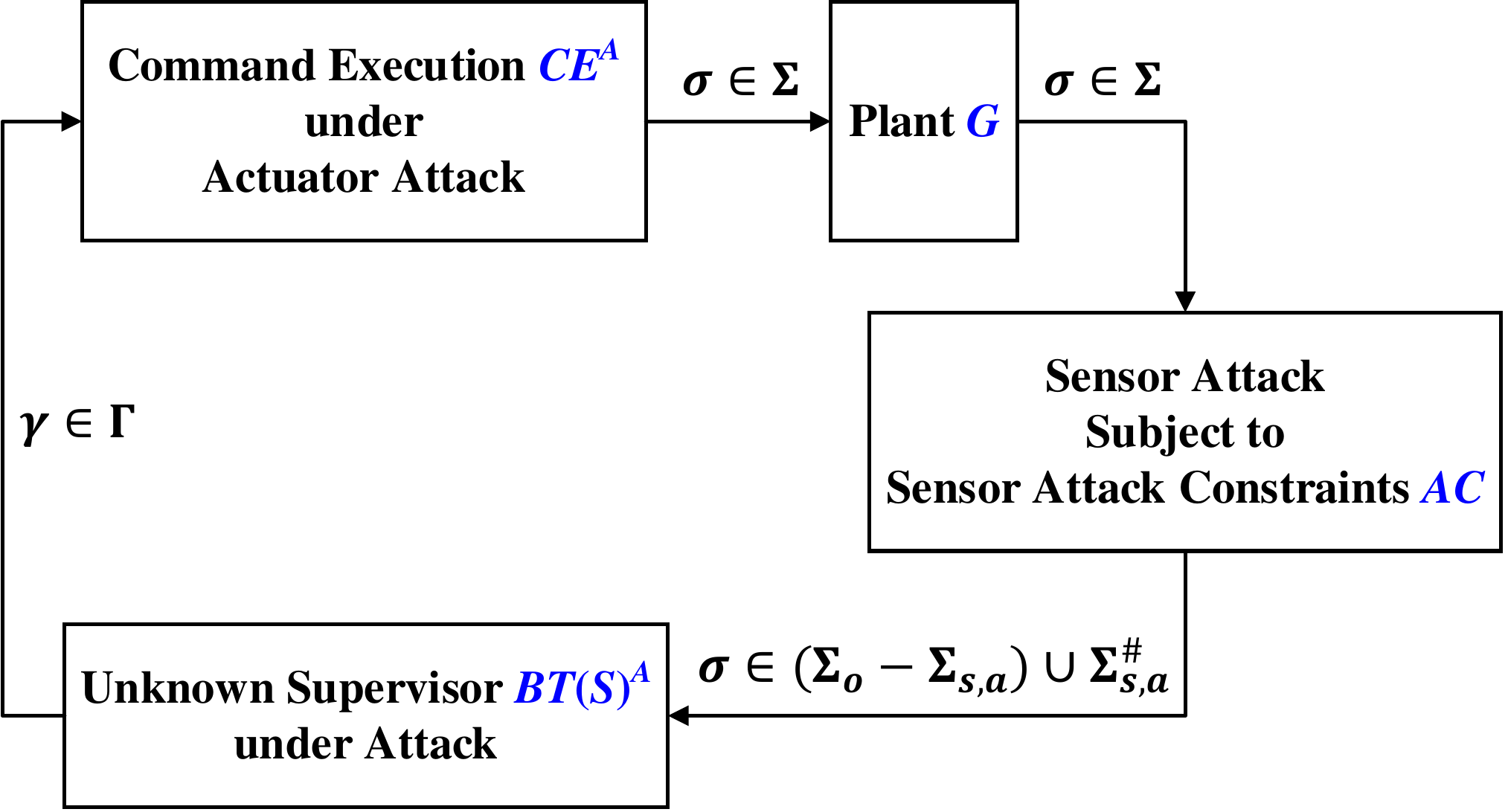}   
\caption{System architecture under sensor-actuator attack}
\label{fig:System architecture under attack}
\end{center}        
\end{figure}

In this section, we shall introduce the system architecture under sensor-actuator attack~\cite{LS20J} and the model of each component. The system architecture is shown in Fig. \ref{fig:System architecture under attack}, which consists of the following components:
\begin{itemize}
\setlength{\itemsep}{3pt}
\setlength{\parsep}{0pt}
\setlength{\parskip}{0pt}
    \item Plant $G$.
    \item Command execution $CE^{A}$ under actuator attack.
    \item Sensor attack subject to sensor attack constraints $AC$.
    \item Unknown supervisor $BT(S)^{A}$ under attack (with an explicit control command sending phase).  
\end{itemize}
In this work, we shall assume that $\Sigma_{c} \subseteq \Sigma_{o}$, i.e., the normality property holds, which is a property that can be easily satisfied in reality. Even if this assumption is relaxed, the proposed synthesis algorithm still is guaranteed to be sound and generates more permissive solutions than the heuristic algorithm proposed in~\cite{LTZS20}, but it is then generally incomplete. For more details, the reader is referred to Remark IV.1. in Section IV. In the following, we explain how the models shown in the system architecture of Fig. 1 can be constructed. %+In any case, it is more permissive than the solution proposed in~\cite{LTZS20}.
% as in real applications, it is typical that all the actuator events are observable to the supervisor. 

\subsection{Sensor attack constraints $AC$}
\label{subsec:sensor attack constraints}
In this work, the basic assumptions of the sensor attacker\footnote{We simply refer to the sensor attack decision making part of the sensor-actuator attacker as the sensor attacker.} is given as follows:
\begin{itemize}
\setlength{\itemsep}{3pt}
\setlength{\parsep}{0pt}
\setlength{\parskip}{0pt}
    \item The sensor attacker can only observe the events in $\Sigma_{o}$, which is the set of observable events of the plant; the set of compromised observable events for the sensor attacker is denoted as $\Sigma_{s,a} \subseteq \Sigma_{o}$.
    \item The sensor attacker can  implement deletion or replacement attacks w.r.t. the events in $\Sigma_{s,a}$. 
    %\item We consider a bounded sensor attack, that is, upon the reception of any event in $\Sigma_{o}$, the number of events that the attack can simultaneously send is upper bounded by $U$, i.e., we consider bounded sensor attacks as in \cite{Su2018}. 
    \item The sensor attack action (deletion or replacement) is instantaneous. When an attack is initiated for a specific observation, it will be completed before the next event can be executed by the plant $G$.
\end{itemize}
%Next, we shall introduce two models: 1) sensor attack constraints, which serves a ``template'' to describe the capabilities of the sensor attack; 2) sensor-actuator attacker, which is the component that we aim to synthesize in this work. 

\begin{figure}[htbp]
\begin{center}

\includegraphics[height=2.4cm]{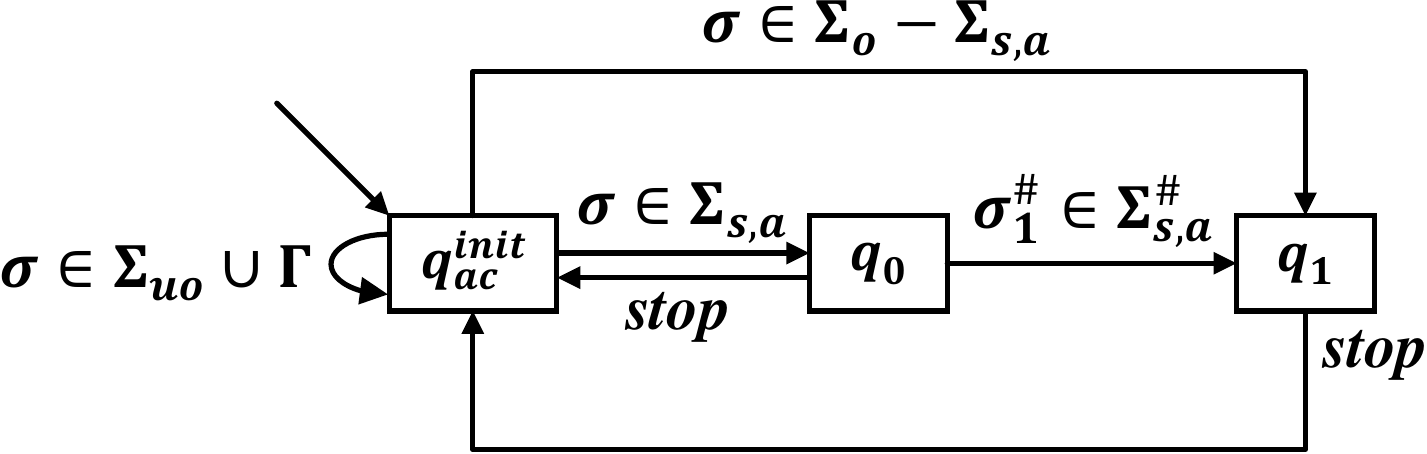}   
\caption{The (schematic) model for sensor attack constraints $AC$}
\label{fig:Sensor attack constraints}
\end{center}        
\end{figure}

%\textbf{Sensor attack constraints:} 
Then, the sensor attack constraints is modelled as a finite state automaton $AC$, shown in Fig. \ref{fig:Sensor attack constraints}.%., which simulates the finite state transducer model of the sensor attack of \cite{Su2018}.
\[
AC = (Q_{ac}, \Sigma_{ac}, \xi_{ac}, q_{ac}^{init})
\]
\begin{itemize}
\setlength{\itemsep}{3pt}
\setlength{\parsep}{0pt}
\setlength{\parskip}{0pt}
    \item $Q_{ac} = \{q_{ac}^{init}, q_{0}, q_{1}\}$
    \item $\Sigma_{ac} = \Sigma \cup \Sigma_{s,a}^{\#} \cup \Gamma \cup \{stop\}$
    \item $\xi_{ac}: Q_{ac} \times \Sigma_{ac} \rightarrow Q_{ac}$
    %\item $Q_{ac,m} = \{q_{ac}^{init}\}$
\end{itemize}
The (partial) transition function $\xi_{ac}$ is defined as follows:
\begin{enumerate}[1.]
\setlength{\itemsep}{3pt}
\setlength{\parsep}{0pt}
\setlength{\parskip}{0pt}
    \item For any $\sigma \in \Sigma_{uo} \cup \Gamma$, $\xi_{ac}(q_{ac}^{init}, \sigma) = q_{ac}^{init}$.
    \item For any $\sigma \in \Sigma_{s,a}$, $\xi_{ac}(q_{ac}^{init}, \sigma) = q_{0}$.
    \item For any $\sigma \in \Sigma_{o} - \Sigma_{s,a}$, $\xi_{ac}(q_{ac}^{init}, \sigma) = q_{1}$.
    \item For any $\sigma \in \Sigma_{s,a}$, $\xi_{ac}(q_{0}, \sigma^{\#}) = q_{1}$.
    \item For any $n \in \{0,1\}$, $\xi_{ac}(q_{n}, stop) = q_{ac}^{init}$.
\end{enumerate}
We shall briefly explain the model $AC$. For the state set, the initial state $q_{ac}^{init}$ denotes that the sensor attacker has not observed any event in $\Sigma_{o}$ since the system initiation or the last attack operation. %$q_{n} (n \in \{0,1\})$ is a state denoting that the sensor attacker has sent $n$ events since the observation of some event in $\Sigma_{o}$. \footnote{In this work, when the sensor attacker observes some event in $\Sigma_{o} - \Sigma_{s,a}$, which cannot be attacked, we shall count such a event in the output of the sensor attack.}
%Thus, we could also interpret those two states as that 
$q_{0}$ ($q_{1}$, respectively) is a state denoting that the sensor attacker has observed some event in $\Sigma_{s,a}$ ($\Sigma_{o} - \Sigma_{s,a}$, respectively).
For the event set, any event $\sigma^{\#}$ in $\Sigma_{s,a}^{\#}$ denotes an event of sending a compromised observable event $\sigma$ to the supervisor by the sensor attacker. Thus, due to the existence of sensor attack, the supervisor can only observe the relabelled copy $\Sigma_{s,a}^{\#}$ instead of $\Sigma_{s,a}$. Any event $\gamma \in \Gamma$ denotes an event of sending a control command $\gamma$ by the supervisor, which will be introduced later in Section \ref{subsec:unknown supervisor}. The event $stop$ denotes the end of the current round of sensor attack operation. In this work, we shall treat any event in $\Sigma_{o} \cup \Sigma_{s,a}^{\#} \cup \{stop\}$ as being observable to the sensor attacker. %{\color{red} theoretically, the transition defined at the initial state labelled by $\sigma \in \Sigma_o-\Sigma_{s, a}$ can also be defined as self-loop instead, having an equal role as $\Gamma$ in command eavesdropping attacker; no need to change}

For the (partial) transition function $\xi_{ac}$, 
\begin{itemize}
\setlength{\itemsep}{3pt}
\setlength{\parsep}{0pt}
\setlength{\parskip}{0pt}
    \item Case 1 says that the occurrence of any event in $\Sigma_{uo} \cup \Gamma$, which is unobservable to the sensor attacker and cannot be attacked, would only lead to a self-loop at the state $q_{ac}^{init}$. The purpose of adding Case 1 is to ensure 1) the alphabet of $AC$ is $\Sigma \cup \Sigma_{s,a}^{\#} \cup \Gamma \cup \{stop\}$, and 2) any event $\sigma \in \Sigma_{uo} \cup \Gamma$ is not defined at non-$q_{ac}^{init}$ states and thus any event in $\Sigma_{o}$ is immediately followed by an event in $\Sigma_{s,a}^{\#} \cup \{stop\}$ to simulate the immediate attack operation or the end of the attack operation following the observation of an event in $\Sigma_{o}$.
    \item Case 2 and Case 3 say that the observation of any event in $\Sigma_{s,a}$ ($\Sigma_{o} - \Sigma_{s,a}$, respectively) would lead to a transition to the state $q_{0}$ ($q_{1}$, respectively), where the sensor attacker may perform some attack operations.
    \item Case 4 says that at the state $q_{0}$, i.e., the sensor attacker has just observed some compromised observable event in $\Sigma_{s,a}$, it can implement sensor replacement attacks by replacing what it observes with any compromised observable event in $\Sigma_{s,a}$. %\footnote{When the sensor attack observes $\sigma \in \Sigma_{s,a}$ and transits to the state $q_{0}$, then it outputs $\sigma^{\#}$ and transits to the state $q_{1}$. Such a case means that the sensor attack actually does not attack the event $\sigma$.}
    \item Case 5 says that at the state $q_{0}$ or $q_{1}$, the sensor attacker can end the current round of sensor attack operation. Indeed, at the state $q_{0}$, such a transition labelled by $stop$ corresponds to the sensor deletion attack. %since for any compromised event in $\Sigma_{s,a}$, the supervisor could only observe its relabelled copy in $\Sigma_{s,a}^{\#}$.
\end{itemize}
Based on the model of $AC$, we know that $|Q_{ac}| = 3$.

\subsection{Plant $G$}
\label{subsec:Plant}

Plant is modelled by a finite state automaton $G = (Q, \Sigma, \xi, q^{init})$. We use $Q_{d} \subseteq Q$ to denote the set of bad states in $G$, any state of which is a goal state that the sensor-actuator attacker targets to induce the plant $G$ to reach. We shall assume each state in $Q_{d}$ is deadlocked, since damage cannot be undone\footnote{Since each state in $Q_d$ is deadlocked, we can also merge these equivalent states into one deadlocked state.}. 

\subsection{Command execution $CE^{A}$ under actuator attack}
\label{subsec:Command execution}

The input to the plant is the control commands in $\Gamma$, while the output of the plant is the events in $\Sigma$. There is thus a ``transduction"  from the input $\gamma \in \Gamma$ of $G$ to the output $\sigma \in \Sigma$ of $G$, which requires an automaton model over $\Sigma \cup \Gamma$ that describes how the control commands are executed in the plant. This automaton model is referred to as the command execution automaton $CE$~\cite{LZS19},~\cite{LS20J},~\cite{zhu2019}. The model of $CE$ is given as follows:
\[
CE = (Q_{ce}, \Sigma_{ce}, \xi_{ce}, q_{ce}^{init})
\]
\begin{itemize}
\setlength{\itemsep}{3pt}
\setlength{\parsep}{0pt}
\setlength{\parskip}{0pt}
    \item $Q_{ce} = \{q^{\gamma}|\gamma \in \Gamma\} \cup \{q_{ce}^{init}\}$
    \item $\Sigma_{ce} = \Gamma \cup \Sigma$
    \item $\xi_{ce}: Q_{ce} \times \Sigma_{ce} \rightarrow Q_{ce}$
\end{itemize}
The (partial) transition function $\xi_{ce}$ is defined as follows:
\begin{enumerate}[1.]
\setlength{\itemsep}{3pt}
\setlength{\parsep}{0pt}
\setlength{\parskip}{0pt}
    \item For any $\gamma \in \Gamma$, $\xi_{ce}(q_{ce}^{init}, \gamma) = q^{\gamma}$.
    \item For any $\sigma \in \gamma \cap \Sigma_{uo}$, $\xi_{ce}(q^{\gamma}, \sigma) = q^{\gamma}$.
    \item For any $\sigma \in \gamma \cap \Sigma_{o}$, $\xi_{ce}(q^{\gamma}, \sigma) = q_{ce}^{init}$.
    %\item For any $\sigma \in \Sigma_{uc}$, $\xi_{ce}(q_{ce}^{init}, \sigma) = q_{ce}^{init}$. we have this transition due to the deletion attack
\end{enumerate}
We shall briefly explain the model $CE$. For the state set, 1) $q_{ce}^{init}$ is the initial state, denoting that the command execution automaton is not using any control command; 2) $q^{\gamma}$ is a state denoting that the command execution automaton is using the control command $\gamma$. 

For the (partial) transition function $\xi_{ce}$, 
\begin{itemize}
\setlength{\itemsep}{3pt}
\setlength{\parsep}{0pt}
\setlength{\parskip}{0pt}
    \item Case 1 says that once $CE$ starts to use $\gamma$, it will transit to the state $q^{\gamma}$.
    \item Cases 2 and 3 say that at the state $q^{\gamma}$, the execution of any event in $\gamma \cap \Sigma_{uo}$ will lead to a self-loop, that is, $\gamma$ will be reused, and the execution of any event in $\gamma \cap \Sigma_{o}$ will lead to the transition to the initial state, that is, $CE$ will wait for the next control command to be issued from the supervisor. 
    %\item With Case 2 and Case 3, Case 4 says that uncontrollable events are always allowed to be fired.
\end{itemize}

\begin{figure}[htbp]
\begin{center}
\includegraphics[height=2.45cm]{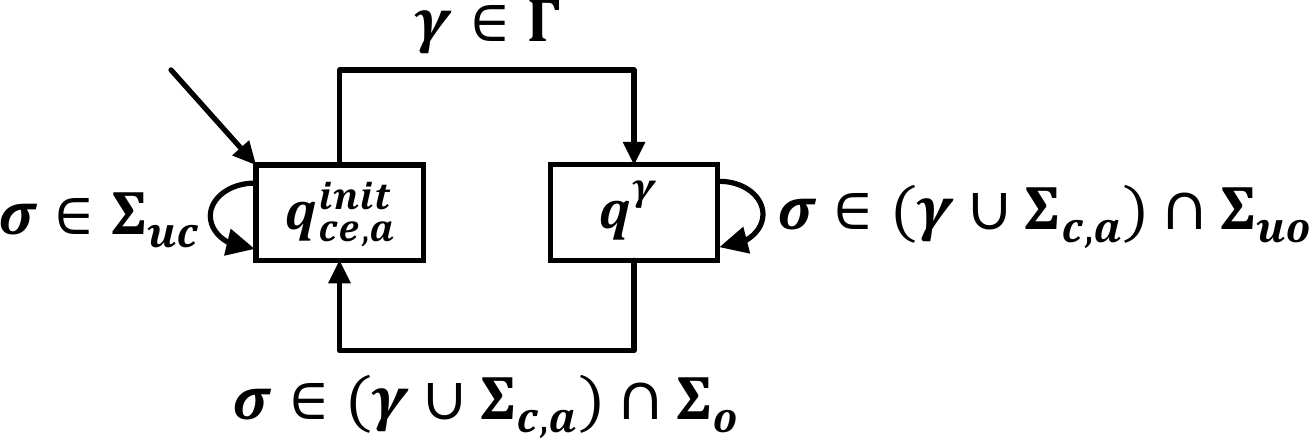}   
\caption{The (schematic) model for command execution automaton $CE^A$ under actuator attack}
\label{fig:Command execution automaton}
\end{center}        
\end{figure}

Next, we shall construct the command execution automaton under actuator attack, denoted as $CE^{A}$. In this work, we consider a class of actuator attackers that can implement both the enablement and disablement attacks, that is, the actuator attacker is capable of modifying the control command $\gamma$ (issued by the supervisor) by enabling or disabling some events in a specified attackable subset $\Sigma_{c,a} \subseteq \Sigma_{c}$, where $\Sigma_{c}$ is the set of controllable events~\cite{LZS19}. Then, based on $CE$, we shall encode the impacts of actuator attack on the event execution phase, and generate the command execution automaton $CE^{A}$ under actuator attack~\cite{LS20},~\cite{LS20J}, which is shown in Fig. \ref{fig:Command execution automaton}. 
\[
CE^{A} = (Q_{ce,a}, \Sigma_{ce,a}, \xi_{ce,a}, q_{ce,a}^{init})
\]
\begin{itemize}
\setlength{\itemsep}{3pt}
\setlength{\parsep}{0pt}
\setlength{\parskip}{0pt}
    \item $Q_{ce,a} = Q_{ce}$
    \item $\Sigma_{ce,a} = \Sigma_{ce} = \Gamma \cup \Sigma$
    \item $\xi_{ce,a}: Q_{ce,a} \times \Sigma_{ce,a} \rightarrow Q_{ce,a}$
    \item $q_{ce,a}^{init} = q_{ce}^{init}$
\end{itemize}
The (partial) transition function $\xi_{ce,a}$ is defined as follows:
\begin{enumerate}[1.]
\setlength{\itemsep}{3pt}
\setlength{\parsep}{0pt}
\setlength{\parskip}{0pt}
    \item For any $q, q' \in Q_{ce,a}$ and any $\sigma \in \Sigma_{ce,a}$, $\xi_{ce}(q, \sigma) = q' \Rightarrow \xi_{ce,a}(q, \sigma) = q'$.
    \item For any $q \in \{q^{\gamma}|\gamma \in \Gamma\}$ and any $\sigma \in \Sigma_{c,a} \cap \Sigma_{uo}$, $\xi_{ce,a}(q, \sigma) = q$.
    \item For any $q \in \{q^{\gamma}|\gamma \in \Gamma\}$ and any $\sigma \in \Sigma_{c,a} \cap \Sigma_{o}$,  $\xi_{ce,a}(q, \sigma) = q_{ce,a}^{init}$.
    \item For any $\sigma \in \Sigma_{uc}$, $\xi_{ce}(q_{ce}^{init}, \sigma) = q_{ce}^{init}$. %{\color{red} this can be moved to case 4 in the definition of $CE$ instead, then case 1 in this definition already takes care of this..; once we consider deletion attacks, can also renew to $\Gamma=2^{\Sigma_c}$, but this will complicate the bipartite supervisor a bit; no need to change..}
\end{enumerate}
In the above definition of $\xi_{ce,a}$, 
\begin{itemize}
\setlength{\itemsep}{3pt}
\setlength{\parsep}{0pt}
\setlength{\parskip}{0pt}
    \item Case 1 retains all the transitions defined in $CE$. 
    \item Due to the existence of  actuator attack, which can enable or disable the events in $\Sigma_{c,a}$, in Case 2 and Case 3 we need to model the occurrences of any attackable event in $\Sigma_{c,a}$, where the execution of an unobservable event in $\Sigma_{c,a} \cap \Sigma_{uo}$ will  lead to a self-loop and the execution of an observable event in $\Sigma_{c,a} \cap \Sigma_{o}$ will lead to the transition back to the initial state $q_{ce,a}^{init}$. 
    \item In Case 4, we need to add the transitions labelled by the uncontrollable events at the initial state $q_{ce}^{init}$ because the sensor attacker considered in this work can carry out sensor deletion attack on some compromised observable event in $\Sigma_{s,a}$, resulting in that the occurrence of this event cannot be observed by the supervisor and thus no control command is issued by the supervisor; in this case, although the command execution automaton receives no control command from the supervisor, it could still execute uncontrollable events, if they are defined at the current state of the plant $G$, since uncontrollable events are always allowed to be fired \footnote{For the model of $CE$, we do not need to add the self-loops labelled by the uncontrollable events at the initial state, since $CE$ describes the execution model in the absence of attack. That is, once the plant fires an observable event, the supervisor will definitely observe the event and immediately issue a control command containing all the uncontrollable events.}.
\end{itemize}
Based on the model of $CE^A$, we know that $|Q_{ce,a}| = |\Gamma| + 1$.

\vspace{0.1cm}

\begin{figure}[htbp]
\begin{center}
\includegraphics[height=5.15cm]{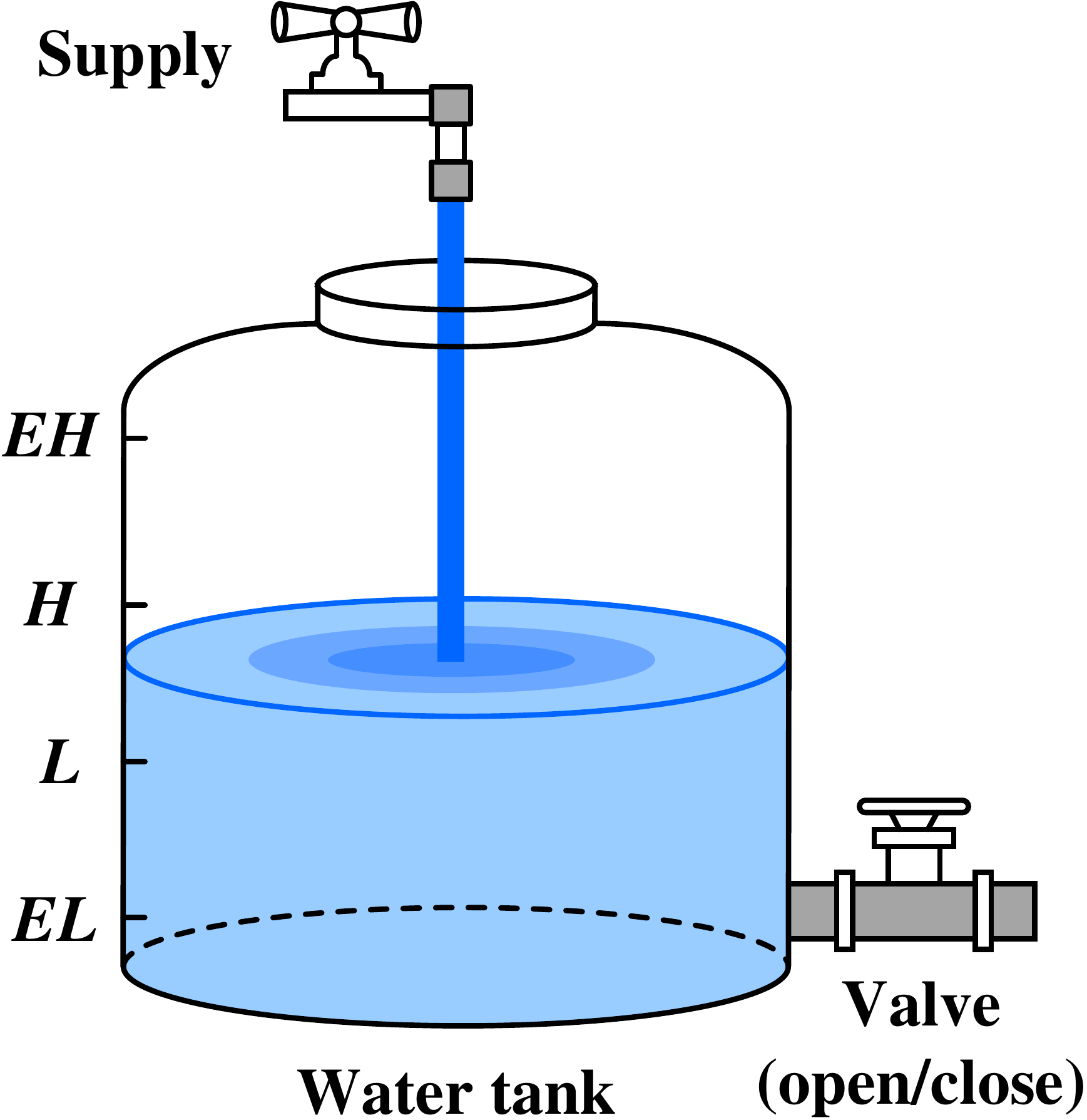}   
\caption{The schematic diagram of the water tank operation scenario}
\label{fig:schematic diagram of the water tank}
\end{center}        
\end{figure}

\begin{figure}[htbp]
\begin{center}
\includegraphics[height=3.2cm]{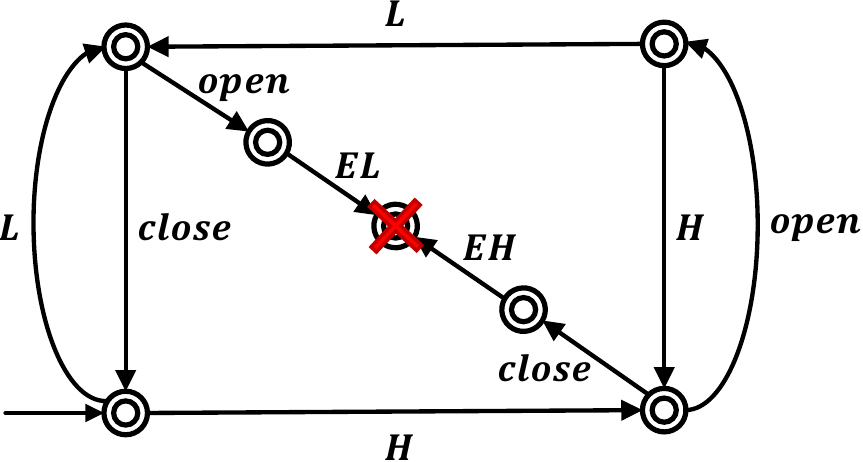}   
\caption{Plant $G$}
\label{fig:Plant G}
\end{center}        
\end{figure}

\begin{figure}[htbp]
\centering
\subfigure[]{
\begin{minipage}[t]{0.43\linewidth}
\centering
\includegraphics[height=1.5in]{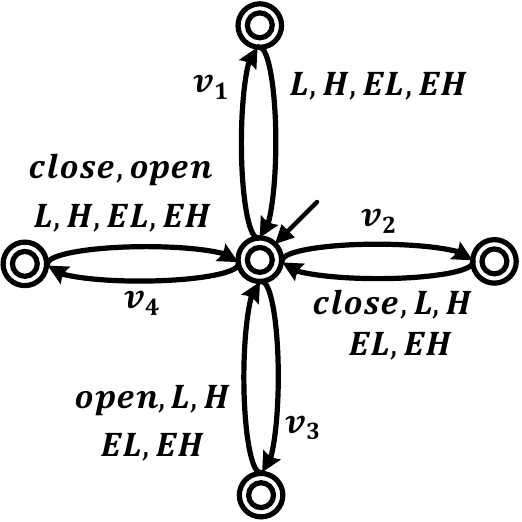}
\end{minipage}
}
\subfigure[]{
\begin{minipage}[t]{0.43\linewidth}
\centering
\includegraphics[height=1.5in]{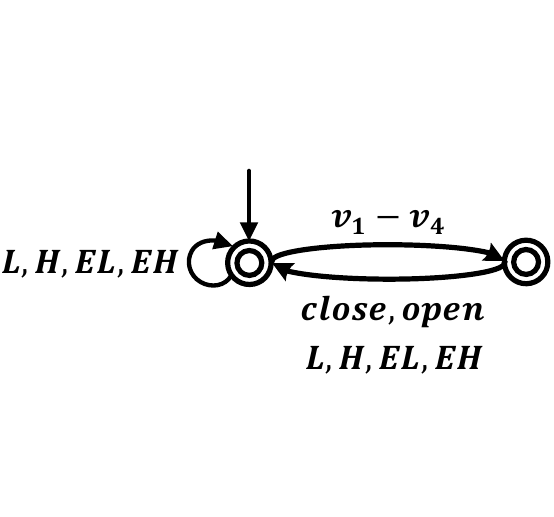}
\end{minipage}
}

\centering
\caption{(a) Command execution automaton $CE$. (b) Command execution automaton $CE^{A}$ under actuator attack (after automaton minimization).}
\label{fig:Example_command execution}
\end{figure}

\begin{figure}[htbp]
\begin{center}
\includegraphics[height=2.4cm]{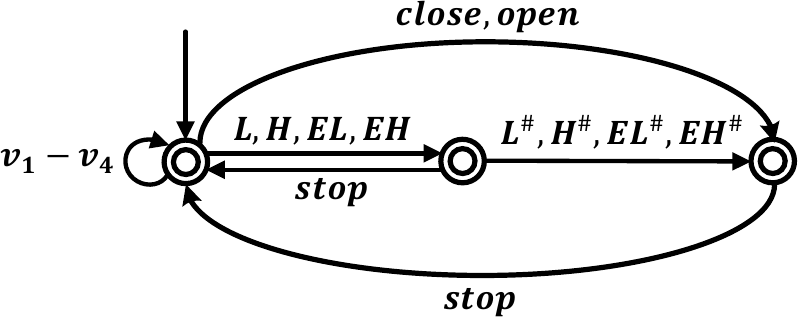}   
\caption{Sensor attack constraints $AC$}
\label{fig:Example_Sensor attack constraints AC}
\end{center}        
\end{figure}

\textbf{Example III.1} We adapt the water tank example from \cite{Su2018} as a running example, whose schematic diagram is shown in Fig. \ref{fig:schematic diagram of the water tank}. The system consists of a constant supply rate, a water tank, and a control valve at the bottom of the tank controlling the outgoing flow rate. We assume the valve can only be fully open or fully closed, resulting in the two events:  $open$ and $close$. The water level can be measured, whose value can trigger some predefined events that denote the water levels: low ($L$), high ($H$), extremely low ($EL$) and extremely high ($EH$). Our control goal is to adjust the control valve operation such that the water level would not be extremely low or extremely high. We assume all the events are observable, i.e., $\Sigma_{o} = \Sigma = \{L, H, EL, EH, close, open\}$. $\Sigma_{c,a} = \Sigma_{c} = \{close, open\}$. $\Sigma_{s,a} = \{L, H, EL, EH\}$. $\Gamma = \{v_{1}, v_{2}, v_{3}, v_{4}\}$. $v_{1} = \{L, H, EL, EH\}$. $v_{2} = \{close, L, H, EL, EH\}$. $v_{3} = \{open, L, H, EL, EH\}$. $v_{4} = \{close, open, L, H, EL, EH\}$. The model of the plant $G$ (the state marked by red cross is the bad state), command execution automaton $CE$, command execution automaton $CE^{A}$ under actuator attack, and the sensor attack constraints $AC$ are shown in Fig. \ref{fig:Plant G} - Fig. \ref{fig:Example_Sensor attack constraints AC}, respectively.

%\begin{figure}[htbp]
%\begin{center}
%\includegraphics[height=4.05cm]{CE.pdf}   
%\caption{Command execution automaton $CE$}
%\label{fig:Command execution automaton CE}
%\end{center}        
%\end{figure}

%\begin{figure}[htb]
%\begin{center}
%\includegraphics[height=1.7cm]{CE_A.pdf}   
%\caption{Command execution automaton $CE^{A}$ under actuator attack}
%\label{fig:Command execution automaton CE_A under actuator attack}
%\end{center}        
%\end{figure}

\subsection{Unknown supervisor $BT(S)^{A}$ under attack}
\label{subsec:unknown supervisor}

%In the Ramadge-Wonham supervisory control theory, under 
In the absence of attacks, a supervisor $S$ over the control constraint $\mathcal{C}=(\Sigma_{c}, \Sigma_{o})$ is often modelled by a finite state automaton $S = (Q_{s}, \Sigma_{s} = \Sigma, \xi_{s}, q_{s}^{init})$, which satisfies the controllability and observability constraints \cite{B1993}:
\begin{itemize}
\setlength{\itemsep}{3pt}
\setlength{\parsep}{0pt}
\setlength{\parskip}{0pt}
    \item (Controllability) For any state $q \in Q_{s}$ and any event $\sigma \in \Sigma_{uc}$, $\xi_{s}(q, \sigma)!$,
    \item (Observability) For any state $q \in Q_{s}$ and any event $\sigma \in \Sigma_{uo}$, if $\xi_{s}(q, \sigma)!$, then $\xi_{s}(q, \sigma) = q$.
\end{itemize}
The control command issued by the supervisor $S$ at state $q \in Q_{s}$ is defined to be $\Gamma(q) = En_{S}(q) = \{\sigma \in \Sigma|\xi_{s}(q,\sigma)!\}$. We assume the supervisor $S$ will immediately issue a control command to the plant whenever an event $\sigma \in \Sigma_{o}$ is received or when the system initiates. 
%In the following text, we shall always consider the supervisors, under which the plant $G$ would not reach any bad state in $Q_{d}$. 

\begin{figure}[htbp]
\begin{center}
\includegraphics[height=2.6cm]{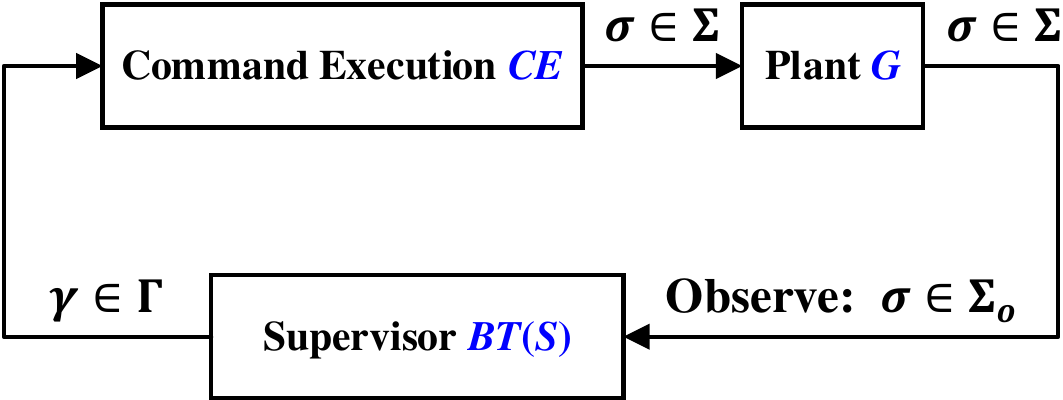} 
\caption{The refined diagram of the supervisory control feedback loop}
\label{fig:Supervisory_Control_Bipartite_Supervisor}
\end{center}        
\end{figure}

Based on the command execution automaton $CE$ and the plant $G$, we shall notice that while $CE$ can model the transduction from $\Gamma$ to $\Sigma$, the transduction needs to be restricted by the behavior of $G$. Thus, only $CE||G$ models the transduction from the input $\Gamma$ of $G$ to the output $\Sigma$ of $G$. The diagram of the supervisory control feedback loop (in the absence of attack) can then be refined as in Fig. \ref{fig:Supervisory_Control_Bipartite_Supervisor} where $BT(S)$, to be introduced shortly, is a control-equivalent bipartite \footnote{Strictly speaking, $BT(S)$ is not bipartite as unobservable events in $\Sigma_{uo}$ would lead to self-loops. In this work, for convenience, we shall always call supervisors with such structures bipartite ones.} supervisor to $S$ and explicitly models the control command sending phase. 

Next, we shall show how to model this bipartite supervisor $BT(S)$ based on $S$~\cite{LZS19}. 
For any supervisor $S = (Q_{s}, \Sigma_{s} = \Sigma, \xi_{s}, q_{s}^{init})$, the procedure to construct $BT(S)$ is given as follows:
\[
BT(S) = (Q_{bs}, \Sigma_{bs}, \xi_{bs}, q_{bs}^{init})
\]
\begin{enumerate}[1.]
\setlength{\itemsep}{3pt}
\setlength{\parsep}{0pt}
\setlength{\parskip}{0pt}
    \item $Q_{bs} = Q_{s} \cup Q_{s}^{com}$, where $Q_{s}^{com}:= \{q^{com} \mid q \in Q_s\}$
    \item $\Sigma_{bs} = \Sigma \cup \Gamma$
    \item \begin{enumerate}[a.]
        \setlength{\itemsep}{3pt}
        \setlength{\parsep}{0pt}
        \setlength{\parskip}{0pt}
            \item $(\forall q^{com} \in Q_{s}^{com}) \, \xi_{bs}(q^{com}, \Gamma(q)) = q$
            \item $(\forall q \in Q_{s})(\forall \sigma \in \Sigma_{uo}) \, \xi_{s}(q, \sigma)! \Rightarrow \xi_{bs}(q, \sigma) = \xi_{s}(q, \sigma) =q$ 
            \item $(\forall q \in Q_{s})(\forall \sigma \in \Sigma_{o}) \, \xi_{s}(q, \sigma)! \Rightarrow \xi_{bs}(q, \sigma) = (\xi_{s}(q, \sigma))^{com}$
        \end{enumerate}
    \item $q_{bs}^{init} = (q_{s}^{init})^{com}$
\end{enumerate}
We shall briefly explain the above construction procedure. For the state set, we add $Q_{s}^{com}$, which is a relabelled copy of $Q_{s}$ with the superscript ``com'' attached to each element of $Q_{s}$. Any state $q^{com} \in Q_{s}^{com}$ is a control state denoting that the supervisor is ready to issue the control command $\Gamma(q)$. Any state $q \in Q_{s}$ is a reaction state denoting that the supervisor is ready to react to an event $\sigma \in \Gamma(q)$.
For the (partial) transition function $\xi_{bs}$, Step 3.a says that at any control state $q^{com} \in Q_{s}^{com}$, after issuing the control command $\Gamma(q)$, the supervisor would transit to the reaction state $q$. Step 3.b says that at any reaction state $q \in Q_{s}$, the occurrence of any unobservable event $\sigma \in \Sigma_{uo}$, if it is defined, would lead to a self-loop, i.e., the state still remains at the reaction state $q$. Step 3.c says that at any reaction state $q \in Q_{s}$, the occurrence of any observable event $\sigma \in \Sigma_{o}$, if it is defined, % such that $\xi_{s}(q,\sigma)!$ 
would lead to a transition to the control state $(\xi_{s}(q, \sigma))^{com}$. The initial state of $BT(S)$ is changed to the initial control state $(q_{s}^{init})^{com}$ which would issue the initial control command $\Gamma(q_{s}^{init})$ when the system initiates. Thus, if we abstract $BT(S)$ by merging the states $x_{com}$ and $x$, treated as equivalent states in the abstraction, then we can recover $S$. In this sense, $BT(S)$ is control equivalent to $S$~\cite{LS20J}. 
%Thus, under this transformed bipartite supervisor $BT(S)$ and the command execution automaton $CE$, the diagram of the supervisory control feedback loop can then be refined as in Fig. \ref{fig:Supervisory_Control_Bipartite_Supervisor}.

In this work, the model of the supervisor is unknown to the adversary, but we assume a safe supervisor has been implemented, that is, in $G||CE||BT(S)$, we assume no plant state in $Q_{d}$ can be reached. Since the attacker can only observe events in $\Sigma_{o}$, the only prior knowledge available to the adversary is the model of the plant $G$ and a set of observations $O \subseteq P_{o}(L(G||CE||BT(S)))$, where $P_{o}: (\Sigma \cup \Gamma)^{*} \rightarrow \Sigma_{o}^{*}$ (or $O \subseteq P_{o}(L(G||S))$, where\footnote{We here abuse the notation $P_o$ for two different natural projections from different domains. But it shall be clear which natural projection we refer to in each case. } $P_{o}: \Sigma^{*} \rightarrow \Sigma_{o}^{*}$) \cite{LTZS20} of the system executions under the unknown supervisor. The set of the attacker's observations $O$ is captured by a finite state automaton $M_{o} = (Q_{o}, \Sigma_{o}, \xi_{o}, q_{o}^{init})$, i.e., $O = L(M_{o})$. We refer to $M_o$ as the observation automaton. Since $O$ is finite, without loss of generality, we assume there is exactly one deadlocked state $q_{o}^{dl} \in Q_{o}$ in $M_{o}$ and, for any maximal string $s \in O$ (in the prefix ordering \cite{WMW10}), we have $\xi_{o}(q_{o}^{init}, s) = q_{o}^{dl}$~\cite{LTZS20}. Then, we have the following definition. 

\emph{Definition III.1 (Consistency)} Given the plant $G$, a supervisor $S$ is said to be consistent with a set of observations $O$ if $O \subseteq P_{o}(L(G||CE||BT(S)))$, where $P_{o}: (\Sigma \cup \Gamma)^{*} \rightarrow \Sigma_{o}^{*}$ (or $O \subseteq P_{o}(L(G||S)$, where $P_{o}: \Sigma^{*} \rightarrow \Sigma_{o}^{*}$).

\vspace{0.1cm}

\begin{figure}[htbp]
\begin{center}
\includegraphics[height=2.15cm]{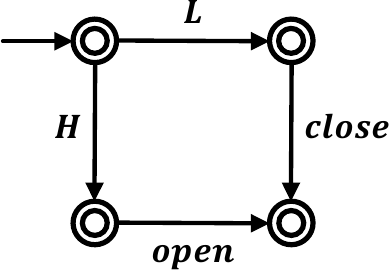}   
\caption{Observations $M_{o}$}
\label{fig:Observations M_o}
\end{center}        
\end{figure}

\textbf{Example III.2} We shall continue with the water tank example. We assume the attacker has collected a set of observations $O$, which is captured by $M_{o}$ shown in Fig. \ref{fig:Observations M_o}. 

\begin{figure}[htbp]
\begin{center}
\includegraphics[height=4cm]{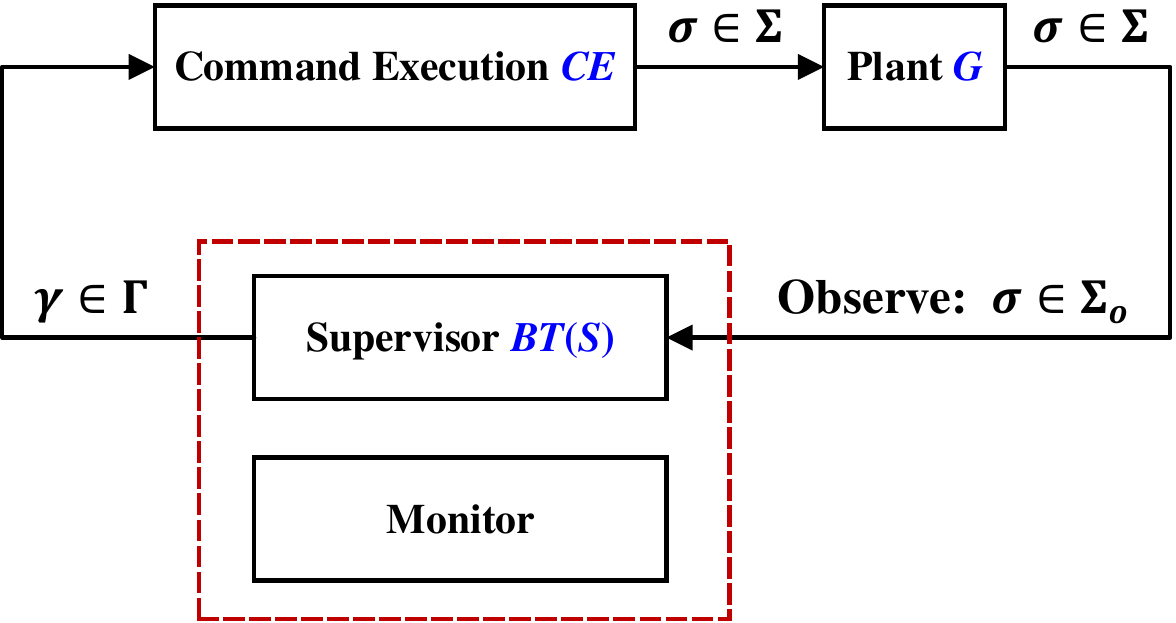} 
\caption{The supervisory control feedback loop with an embedded monitor}
\label{fig:Monitor_Embedding_Reasoning}
\end{center}        
\end{figure}

In this work, since we take the attack into consideration  and aim to synthesize a covert sensor-actuator attacker against the unknown supervisor, we shall modify $BT(S)$ to generate a new bipartite supervisor $BT(S)^{A}$ under attack by modelling the effects of the sensor-actuator attack on the supervisor. The construction of $BT(S)^{A}$ consists of the following steps, including \textbf{Step 1} and \textbf{Step 2}:

\textbf{Step 1:} Firstly, in this work, we assume the monitoring \cite{LS20} function is embedded into the supervisor, that is, the supervisor is able to compare its online observations of the system execution with the ones that can be observed in the absence of attack, and once some information inconsistency happens, the supervisor can assert the existence of an attacker and halts the system operation. To embed the monitoring mechanism into the supervisor in the absence of attack, we adopt what we refer to as the universal monitor $P_{\Sigma_{o} \cup \Gamma}(G||CE)$ to refine $BT(S)$ by synchronous product and obtain $BT(S)||P_{\Sigma_{o} \cup \Gamma}(G||CE)$.
To see that $P_{\Sigma_{o} \cup \Gamma}(G||CE)$ is a universal monitor that works for any supervisor $S$, we perform the diagrammatic reasoning as follows (see Fig. \ref{fig:Monitor_Embedding_Reasoning}). 

%{\color{red} Here, may need to draw a picture that adds the monitor that sits below supervisor and looks into $\Gamma$ and $\Sigma_o$, we can discuss on this.. to let the user immediately see our reasoning.}
\begin{enumerate}[1.]
    \item Since the monitor observes the output and the input of the supervisor $S$, it can observe the events in $\Sigma_o \cup \Gamma$. 
    \item The model of the universal monitor is then exactly its observable model $P_{\Sigma_o \cup \Gamma}(G || CE)$ of everything that is external to $S$, i.e., $G || CE$.
\end{enumerate}
We refer to $P_{\Sigma_o \cup \Gamma}(G || CE)$ as a universal monitor as it looks external against $S$ and thus ignores the model of $S$; intuitively, it is a monitor that works for any supervisor $S$. Then,
when the monitoring mechanism is embedded into the supervisor, we simply refine the universal monitor $P_{\Sigma_o \cup \Gamma}(G || CE)$ with the supervisor model $BT(S)$ to obtain $BT(S)||P_{\Sigma_{o} \cup \Gamma}(G||CE)$. We record this automaton as $BT(S)^1$. Thus, 
\[
BT(S)^{1} = BT(S)||P_{\Sigma_{o} \cup \Gamma}(G||CE) = (Q_{bs,1}, \Sigma \cup \Gamma, \xi_{bs,1}, q_{bs,1}^{init})
\]
It is noteworthy that $BT(S)^{1}$ is bipartite as $BT(S)$ is bipartite. Thus, we could partition the state set $Q_{bs,1}$ into two parts, $Q_{bs,1} = Q_{bs,1}^{rea} \cup Q_{bs,1}^{com}$, where at any state of $Q_{bs,1}^{rea}$, only events in $\Sigma$ are defined, and at any state of $Q_{bs,1}^{com}$, only events in $\Gamma$ are defined. Then, we write
\[
BT(S)^{1} = (Q_{bs,1}^{rea} \cup Q_{bs,1}^{com}, \Sigma \cup \Gamma, \xi_{bs,1}, q_{bs,1}^{init})
\]
We have the following useful results. 

\vspace{0.1cm}

\emph{Proposition III.1.} $L(BT(S)^{1}||G||CE) = L(BT(S)||G||CE)$.

\emph{Proof:} LHS = $L(BT(S)||P_{\Sigma_{o} \cup \Gamma}(G||CE)||G||CE) = L(BT(S)||G||CE)$ = RHS. \hfill $\blacksquare$

\vspace{0.1cm}

%Firstly, we prove that $L(BT(S)^{1}||G||CE) \subseteq L(BT(S)||G||CE)$, i.e., we need to prove that for any $s \in L(BT(S)^{1}||G||CE)$, we have $s \in L(BT(S)||G||CE)$. Since $s \in L(BT(S)^{1}||G||CE) = L(BT(S)^{1}) \cap P_{G}^{-1}(L(G)) \cap L(CE)$, where $P_{G}: (\Sigma \cup \Gamma)^{*} \rightarrow \Sigma^{*}$, we have $s \in L(BT(S)^{1})$, $s \in P_{G}^{-1}(L(G))$, and $s \in L(CE)$. Based on the construction of $L(BT(S)^{1})$, we know that $s \in L(BT(S))$, thus, $s \in L(BT(S)) \cap P_{G}^{-1}(L(G)) \cap L(CE) = L(BT(S)||G||CE)$

%Secondly, we prove that $L(BT(S)||G||CE) \subseteq L(BT(S)^{1}\\ ||G||CE)$, i.e., we need to prove that for any $s \in L(BT(S)||G||CE)$, we have $s \in L(BT(S)^{1}||G||CE)$. We adopt the contradiction and assume that $s \notin L(BT(S)^{1}||G||CE)$. Since $s \in L(BT(S)||G||CE) = L(BT(S)) \cap P_{G}^{-1}(L(G)) \cap L(CE)$, we have $s \in L(BT(S))$, $s \in P_{G}^{-1}(L(G))$, and $s \in L(CE)$. In addition, since $s \notin L(BT(S)^{1}||G||CE) = L(BT(S)^{1}) \cap P_{G}^{-1}(L(G)) \cap L(CE)$, we have $s \notin L(BT(S)^{1})$. Based on the construction of $BT(S)^{1}$, we know that $L(BT(S)||G||CE) \subseteq L(BT(S)^{1})$, implying that $s \notin L(BT(S)||G||CE)$, which causes the contradiction. Thus, $s \in L(BT(S)^{1}||G||CE)$, implying that $L(BT(S)||G||CE) \subseteq L(BT(S)^{1}||G||CE)$. Then the proof is completed. {\color{red} can use the result that $BT(S)^1 = BT(S) || P_{\Sigma_o \cup \Gamma}(G || CE)$ to have shorter proof.} \hfill $\blacksquare$

\vspace{0.1cm}

\emph{Corollary III.1.} $O \subseteq P_{o}(L(BT(S)^{1}||G||CE))$.

\emph{Proof:} This directly follows from  \emph{Proposition III.1} and the fact that $O \subseteq P_{o}(L(BT(S)||G||CE))$. \hfill $\blacksquare$

\vspace{0.1cm}

%{\color{red} for the following, since $BT(S)^1$ is said to be control equivalent to $BT(S)$, it is viewed as  a supervisor. Automaton equality needs to be defined based on language equality in the Preliminary, def: marked behavior and closed behavior must all be equal for automata to be equal. In step 2, there should be another aspect that says how attacker can be discovered..}
%Based on \emph{Proposition III.1}, $BT(S)^{1}$ is control-equivalent to $BT(S)$. 
%Indeed, it can be checked that $BT(S)||M = BT(S)||P_{\Sigma_{o} \cup \Gamma}(G||CE||BT(S)) = BT(S)||P_{\Sigma_{o} \cup \Gamma}(G||CE)=BT(S)^1$. {\color{red} may need to double check the second equality..it shall be true}

\vspace{0.1cm}
\begin{comment}
We now bring in the concept of the $\mathcal{C}$-abstraction of $G$~\cite{LS20BJ}, where $\mathcal{C}=(\Sigma_{c}, \Sigma_{o})$ is the control constraint for the supervisor $S$. The $\mathcal{C}$-abstraction of $G$ is defined to be 
$P_{\mathcal{C}}(G)=(2^Q, \Sigma \cup \Gamma, \Delta_{\mathcal{C}}, \{q_0\})$, where  $\Delta_{\mathcal{C}}$ is defined as follows.
\begin{enumerate}
    \item for any $\varnothing \neq V \subseteq Q$ and any $\gamma \in \Gamma$, $\Delta_{\mathcal{C}}(V, \gamma)=UR_{G, \gamma \cap \Sigma_{uo}}(V)$
    \item for any $\varnothing \neq V \subseteq Q$ and any $\sigma \in \Sigma_o$, $\Delta_{\mathcal{C}}(V, \sigma)=\delta(V, \sigma)$
    \item for any $\varnothing \neq V \subseteq Q$ and any $\sigma \in \Sigma_{uo}$, $\Delta_{\mathcal{C}}(V, \sigma)=V$
\end{enumerate}
\end{comment}
\emph{Proposition III.2.} $BT(S)||P_{\Sigma_{o} \cup \Gamma}(G||CE||BT(S)) = BT(S)^1$.

\emph{Proof:} It is clear that LHS = $BT(S)||P_{\Sigma_{o} \cup \Gamma}(G||CE||BT(S)) \sqsubseteq  BT(S)||P_{\Sigma_{o} \cup \Gamma}(G||CE)$ = RHS. We have RHS $= BT(S)||P_{\Sigma_{o} \cup \Gamma}(G||CE)  \sqsubseteq BT(S)||P_{\Sigma_{o} \cup \Gamma}(G||CE||BT(S))$=LHS, as the sequences in $P_{\Sigma_{o} \cup \Gamma}(G||CE)$ that can survive the synchronous product with $BT(S)$ must come from %\footnote{Alternatively, we can carry out a proof by contradiction to show that $BT(S)||P_{\Sigma_{o} \cup \Gamma}(G||CE)  \sqsubseteq BT(S)||P_{\Sigma_{o} \cup \Gamma}(G||CE||BT(S))$ or use the notion of $\mathcal{C}$-abstraction $P_{\mathcal{C}}(G)$~\cite{LS20BJ} of $G$ and establish that $BT(S)||P_{\Sigma_{o} \cup \Gamma}(G||CE) \sqsubseteq  BT(S)||P_{\mathcal{C}}(G) = BT(S)||P_{\Sigma_{o} \cup \Gamma}(G||BT(S))= BT(S)||P_{\Sigma_{o} \cup \Gamma}(G||CE||BT(S))$. {\color{red} is this correct? XXXXXXXXXXXXXXXXXXXXXXXXXXXXXXXXXXXXXXXx, if too hard to verify, then we may just remove this...}}
$P_{\Sigma_{o} \cup \Gamma}(G||CE||BT(S))$.\hfill $\blacksquare$

\vspace{0.1cm}

%{\color{red} seem can bring in the $C$-abstraction to prove the right hand side is subset of left}
%{\color{red}for the second equality: left as an inclusion of right is direct, right as inclusion of left (is also correct based on intuition), }

%Intuitively, Proposition III.2 states that $BT(S)^1$ does embed the monitoring mechanism into the supervisor $BT(S)$.

Based on \emph{Proposition III.2}, $BT(S)^{1}$ indeed embeds the monitor $M = P_{\Sigma_{o} \cup \Gamma}(G||CE||BT(S))$ \cite{LS20}, which is adopted to detect the attacker by comparing the online observations with the ones that can be observed in the absence of attack.

\textbf{Step 2:} We shall encode the effects of the sensor-actuator attack into $BT(S)^{1}$ to generate $BT(S)^{A}$. The effects of the sensor-actuator attack include the following: 1) Due to the existence of the sensor attack, for any event $\sigma \in \Sigma_{s,a}$, the supervisor cannot observe it but can observe the relabelled copy $\sigma^{\#} \in \Sigma_{s,a}^{\#}$ instead, 2) Any event in $\Sigma_{c,a} \cap \Sigma_{uo}$ might be enabled by the actuator attack and its occurrence is unobservable to the supervisor, and 3) the covertness-breaking situations can happen, i.e., information inconsistency between the online observations and the ones that can be observed in the absence of attack can happen. The construction procedure of $BT(S)^{A}$ is given as follows:
\[
BT(S)^{A} = (Q_{bs,a}, \Sigma_{bs,a}, \xi_{bs,a}, q_{bs,a}^{init})
\]
\begin{enumerate}[1.]
\setlength{\itemsep}{3pt}
\setlength{\parsep}{0pt}
\setlength{\parskip}{0pt}
    \item $Q_{bs,a} = Q_{bs,1} \cup \{q^{detect}\}= Q_{bs,1}^{rea} \cup Q_{bs,1}^{com} \cup \{q^{detect}\}$
    \item $\Sigma_{bs,a} = \Sigma \cup \Sigma_{s,a}^{\#} \cup \Gamma$
    \item \begin{enumerate}[a.]
        \setlength{\itemsep}{3pt}
        \setlength{\parsep}{0pt}
        \setlength{\parskip}{0pt}
        \item $(\forall q, q' \in Q_{bs,1})(\forall \sigma \in \Sigma_{s,a}) \, \xi_{bs,1}(q, \sigma) = q' \Rightarrow \xi_{bs,a}(q, \sigma^{\#}) = q' \wedge \xi_{bs,a}(q, \sigma) = q$
        \item $(\forall q \in Q_{bs,1}^{rea}) (\forall \sigma \in \Sigma_{c,a} \cap (\Sigma_{uo} \cup \Sigma_{s,a})) \, \xi_{bs,a}(q, \sigma) = q$
        \item $(\forall q, q' \in Q_{bs,1})(\forall \sigma \in (\Sigma - \Sigma_{s,a}) \cup \Gamma) \, \xi_{bs,1}(q, \sigma) = q' \Rightarrow \xi_{bs,a}(q, \sigma) = q'$
        \item $(\forall q \in Q_{bs,1}^{rea})(\forall \sigma \in \Sigma_{o} - \Sigma_{s,a}) \, \neg \xi_{bs,1}(q, \sigma)! \Rightarrow \xi_{bs,a}(q, \sigma) = q^{detect}$
        \item $(\forall q \in Q_{bs,1}^{rea})(\forall \sigma \in \Sigma_{s,a}) \, \neg \xi_{bs,1}(q, \sigma)! \Rightarrow \xi_{bs,a}(q, \sigma^{\#}) = q^{detect}$
    \end{enumerate}
    \item $q_{bs,a}^{init} = q_{bs,1}^{init}$
\end{enumerate}
We shall briefly explain the above procedure for constructing $BT(S)^{A}$. Firstly, at Step 1, all the states in $BT(S)^{1}$ are retained, and we add a new state $q^{detect}$ into the state set to explicitly model that the presence of the attacker is detected. Then, for the (partial) transition function $\xi_{bs,a}$, at Step 3.a, we perform the following: 1) all the transitions labelled by events in $\Sigma_{s,a}$ are replaced with the copies in $\Sigma_{s,a}^{\#}$, denoted by $\xi_{bs,a}(q, \sigma^{\#}) = q{'}$, and 2) the transitions labelled by events in $\Sigma_{s,a}$ and originally defined in $BT(S)^{1}$ at state $q$ would become self-loops since these events can be fired and are unobservable to the supervisor, denoted by $\xi_{bs,a}(q, \sigma) = q$. At Step 3.b, at any reaction state $q \in Q_{bs,1}^{rea}$, we shall add self-loop transitions labelled by the events in $\Sigma_{c,a} \cap (\Sigma_{uo} \cup \Sigma_{s,a})$ since such events can be enabled due to the actuator attack at the state $q$ and are unobservable to the supervisor. At Step 3.c, all the other transitions, labelled by events in $(\Sigma - \Sigma_{s,a}) \cup \Gamma$, in $BT(S)^{1}$ are retained. Step 3.d and Step 3.e are defined to encode the covertness-breaking situations: at any reaction state $q \in Q_{bs,1}^{rea}$, for any observable event $\sigma \in \Sigma_{o}$, we add the transition, labelled by $\sigma \in \Sigma_{o} - \Sigma_{s,a}$ or the relabelled copy $\sigma^{\#} \in \Sigma_{s,a}^{\#}$, to the state $q^{detect}$ if $\neg \xi_{bs,1}(q, \sigma)!$. Intuitively, the event $\sigma$ should not be observed at the state $q$ in the absence of attack. 

Based on the model of $BT(S)^{A}$, we know that $|Q_{bs,a}| \leq 2|Q_{s}| + 1$.

\begin{figure}[htbp]
\begin{center}

\includegraphics[height=4.7cm]{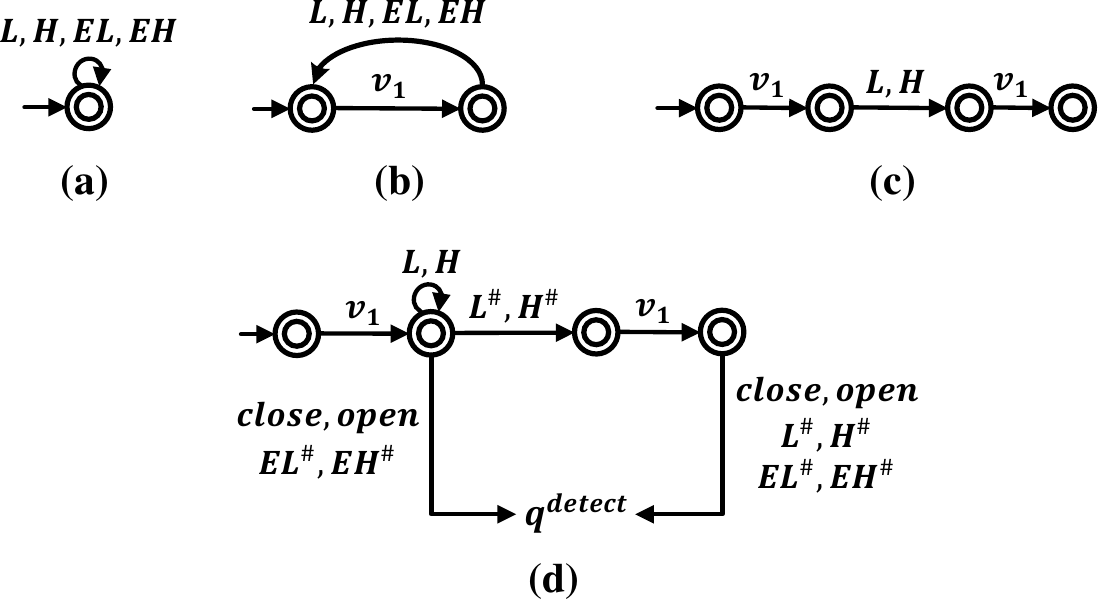}   
\caption{(a) $S$. (b) $BT(S)$. (c) $BT(S)^{1}$. (d) $BT(S)^{A}$.}
\label{fig:Example_S_To_BT(S)A}
\end{center}        
\end{figure}

\textbf{Example III.3} We shall continue with the water tank example. For a supervisor $S$ shown in Fig. \ref{fig:Example_S_To_BT(S)A}. (a), the step-by-step constructed $BT(S)$, $BT(S)^{1}$, and $BT(S)^{A}$ are illustrated in Fig. \ref{fig:Example_S_To_BT(S)A}. (b), (c), (d), respectively.

\subsection{Sensor-actuator attacker}
\label{subsec:sensor-actuator attack}

%\textbf{Sensor-actuator attack:} 
The sensor-actuator attacker is modelled by a finite state automaton $A = (Q_{a}, \Sigma_{a}, \xi_{a}, q_{a}^{init})$, where $\Sigma_{a} = \Sigma \cup \Sigma_{s,a}^{\#} \cup \Gamma \cup \{stop\}$. In addition, there are two conditions that need to be satisfied:
\begin{itemize}
\setlength{\itemsep}{3pt}
\setlength{\parsep}{0pt}
\setlength{\parskip}{0pt}
    \item (A-controllability) For any state $q \in Q_{a}$ and any event $\sigma \in \Sigma_{a,uc} := \Sigma_{a} - (\Sigma_{c,a} \cup \Sigma_{s,a}^{\#} \cup \{stop\})$, $\xi_{a}(q, \sigma)!$ 
    \item (A-observability) For any state $q \in Q_{a}$ and any event $\sigma \in \Sigma_{a,uo} := \Sigma_{a} - (\Sigma_{o} \cup \Sigma_{s,a}^{\#} \cup \{stop\})$, if $\xi_{a}(q, \sigma)$!, then $\xi_{a}(q, \sigma) = q$.
\end{itemize}
A-controllability states that the sensor-actuator attacker can only disable events in $\Sigma_{c,a} \cup \Sigma_{s,a}^{\#} \cup \{stop\}$. A-observability states that the sensor-actuator attacker can only make a state change after observing an event %\footnote{In this work, we shall treat the events in $\Sigma_{s,a}^{\#} \cup \{stop\}$ as being observable to the sensor-actuator attack, although the opposite scenario can also be dealt with in a similar way.}
in $\Sigma_{o} \cup \Sigma_{s,a}^{\#} \cup \{stop\}$. 
%In this work, by construction, all the controllable events for the sensor-actuator attack are also observable to the sensor-actuator attack. 
In the following text, we shall refer to 
\[
\begin{aligned}
\mathscr{C}_{ac} = (\Sigma_{c,a} \cup \Sigma_{s,a}^{\#} \cup \{stop\}, \Sigma_{o} \cup \Sigma_{s,a}^{\#} \cup \{stop\})
\end{aligned}
\] 
as the attacker's control constraint, and $(\Sigma_{o}, \Sigma_{s,a}, \Sigma_{c,a})$ as the attack constraint.
It is apparent that the attacker's control constraint $\mathscr{C}_{ac}$ is uniquely determined by the attack constraint $(\Sigma_{o}, \Sigma_{s,a}, \Sigma_{c,a})$.
%For the command eavesdropping sensor-actuator attacker, we only need to add $\Gamma$ into the set of observable events of the sensor-actuator attacker, which contains the following two modifications: 
%\begin{enumerate}[1.]
%\setlength{\itemsep}{3pt}
%\setlength{\parsep}{0pt}
%\setlength{\parskip}{0pt}
%    \item (A-observability) For any state $q \in Q_{a}$ and any event $\sigma \in \Sigma_{a,uo} = \Sigma_{a} - (\Sigma_{o} \cup \Sigma_{s,a}^{\#} \cup \{stop\} \cup \Gamma)$, if $\xi_{a}(q, \sigma)$!, then $\xi_{a}(q, \sigma) = q$.
%    \item For the sensor-actuator attack-control constraint, $\mathscr{C}_{ac} = (\Sigma_{c,a} \cup \Sigma_{s,a}^{\#} \cup \{stop\}, \Sigma_{o} \cup \Sigma_{s,a}^{\#} \cup \{stop\} \cup \Gamma)$
%\end{enumerate}

%%%%%%%%%%%%%%%%%%%%%%%%%%%%%%%%%%%%%%%%%%%%%%%%%%%%%%%%%%%%%%%%%%%%%%%%%%%%%%%%%%%%

\section{Synthesis of Maximally Permissive Covert Attackers Against Unknown Supervisors}
\label{sec:Synthesis of Maximally Permissive Covert Attackers Against Unknown Supervisors}

In this section, we shall present the solution methodology for the synthesis of maximally permissive covert sensor-actuator attackers against unknown (safe) supervisors by using observations. Firstly, in this work, since we focus on the synthesis of sensor-actuator attacker that aims to cause damage-infliction, we shall denote the marker state set of $G$ as $Q_{d}$, and still denote the modified automaton as $G = (Q, \Sigma, \xi, q^{init}, Q_{d})$ in the following text. Then, based on the above-constructed component models in Section \ref{sec:Component models under sensor-actuator attack}, we know that, given any plant $G$, command execution automaton $CE^{A}$ under actuator attack, sensor attack constraints $AC$, sensor-actuator attacker $A$, bipartite supervisor $BT(S)^{A}$ under attack\footnote{To be precise, $BT(S)^{A}$ is not an attacked supervisor as it  also embeds the model of the attacked monitor.}, the closed-loop behavior is (cf. Fig. 1)
\[
\begin{aligned}
\mathcal{B} = G||CE^{A}||AC||BT(S)^{A}||A = (Q_{b}, \Sigma_{b}, \xi_{b}, q_{b}^{init}, Q_{b,m})
\end{aligned}
\]
\begin{itemize}
\setlength{\itemsep}{3pt}
\setlength{\parsep}{0pt}
\setlength{\parskip}{0pt}
    \item $Q_{b} = Q \times Q_{ce,a} \times Q_{ac} \times Q_{bs,a} \times Q_{a}$
    \item $\Sigma_{b} = \Sigma \cup \Sigma_{s,a}^{\#} \cup \Gamma \cup \{stop\}$
    \item $\xi_{b}: Q_{b} \times \Sigma_{b} \rightarrow Q_{b}$
    \item $q_{b}^{init} = (q^{init}, q_{ce,a}^{init}, q_{ac}^{init}, q_{bs,a}^{init}, q_{a}^{init})$
    \item $Q_{b,m} = Q_{d} \times Q_{ce,a} \times Q_{ac} \times Q_{bs,a} \times Q_{a}$
\end{itemize}
Then, we have the following definitions~\cite{LS20J}.

\vspace{0.1cm}

\emph{Definition IV.1 (Covertness)} Given any plant $G$, command execution automaton $CE^{A}$ under actuator attack, sensor attack constraints $AC$, and bipartite supervisor $BT(S)^A$ under attack, the sensor-actuator attacker $A$ is said to be covert against the supervisor $S$ w.r.t. the  attack constraint $(\Sigma_{o}, \Sigma_{s,a}, \Sigma_{c,a})$ if any state in 
\[
\begin{aligned}
Q_{bad} = \{(q, q_{ce,a}, q_{ac}, q_{bs,a}, q_{a}) \in Q_{b}|q \notin Q_{d} \wedge q_{bs,a} = q^{detect}\}
\end{aligned}
\]
is not reachable in $\mathcal{B}$.

\vspace{0.1cm}

\emph{Definition IV.2 (Damage-reachable)} Given any plant $G$, command execution automaton $CE^{A}$ under actuator attack, sensor attack constraints $AC$, and bipartite supervisor $BT(S)^A$ under attack, the sensor-actuator attacker $A$ is said to be damage-reachable against the supervisor $S$ w.r.t. the attack constraint $(\Sigma_{o}, \Sigma_{s,a}, \Sigma_{c,a})$ if some marker state in $Q_{b,m}$ is reachable in $\mathcal{B}$, that is, $L_{m}(\mathcal{B}) \neq \varnothing$.

\vspace{0.1cm}

\emph{Definition IV.3 (Successful)} Given any plant $G$, a set of observations $O$, and the attack constraint $(\Sigma_{o}, \Sigma_{s,a}, \Sigma_{c,a})$, a sensor-actuator attacker $A$ is said to be successful if it is covert and damage-reachable against any safe supervisor that is consistent with $O$.

\vspace{0.1cm}

\emph{Definition IV.4 (Maximal Permissiveness)} Given any plant $G$, a set of observations $O$, and the attack constraint $(\Sigma_{o}, \Sigma_{s,a}, \Sigma_{c,a})$, a successful sensor-actuator attacker $A$ is said to be maximally permissive if for any other successful sensor-actuator attacker $A'$, we have $L(G||CE^{A}||AC||BT(S)^{A}||A') \subseteq L(G||CE^{A}||AC||BT(S)^{A}||A)$, for any safe supervisor $S$ that is consistent with $O$.

\vspace{0.1cm}

\emph{Remark IV.1} If $L(G||CE^{A}||AC||BT(S)^{A}||A') \subseteq L(G||CE^{A}||AC||BT(S)^{A}||A)$, then we have 
\[
\begin{aligned}
& L_{m}(G||CE^{A}||AC||BT(S)^{A}||A') \\= & L(G||CE^{A}||AC||BT(S)^{A}||A')||L_{m}(G) \\ \subseteq & L(G||CE^{A}||AC||BT(S)^{A}||A)||L_{m}(G) \\ =  & L_{m}(G||CE^{A}||AC||BT(S)^{A}||A)
\end{aligned}
\]

Based on the above definitions, the  observation-assisted covert attacker synthesis problem to be solved in this work is formulated as follows:
 
\vspace{0.1cm}

\textbf{Problem 1:} Given the plant $G$, a set of observations $O$ and the attack constraint $(\Sigma_{o}, \Sigma_{s,a}, \Sigma_{c,a})$, synthesize a maximally permissive successful, i.e., covert and damage-reachable, sensor-actuator attacker?

Next, we shall present our solution methodology for \textbf{Problem 1}.

\subsection{Main idea}
\label{subsec:Main idea of the Solution Methodology}

\begin{figure}[htbp]
\begin{center}

\includegraphics[height=5.7cm]{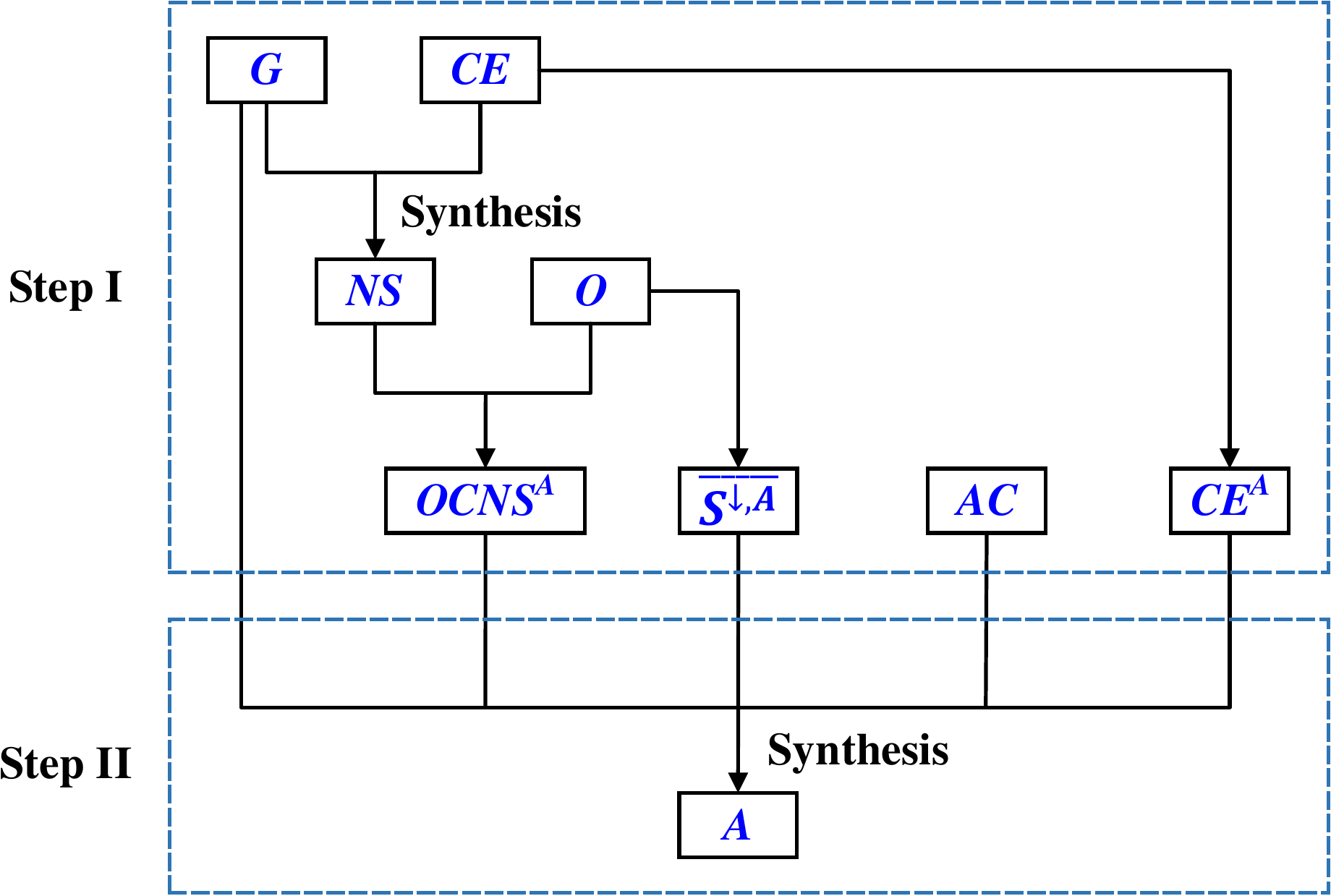}   
\caption{The procedure of the proposed solution methodology}
\label{fig:The procedure of solution methodology}
\end{center}        
\end{figure}

Before we delve into the detailed solution methodology which will be presented in Section \ref{subsec:Solution methodology}, the high-level idea of our method shown in Fig. \ref{fig:The procedure of solution methodology} is explained in the following. We perform the chaining of two synthesis constructions, in order to synthesize maximally permissive covert  damage-reachable sensor-actuator attackers, assisted with the finite set of observations $O$.
\begin{enumerate}[1.]
\item At the first step, based on the plant $G$ and the command execution automaton $CE$, we shall synthesize $NS$, the supremal safe command non-deterministic supervisor~\cite{zhu2019, Linnetworked} that embeds all the safe partial-observation supervisors in the sense of~\cite{YL16}, by using the normality property based synthesis~\cite{WMW10, zhu2019, Linnetworked, WLLW18}. 
%The central ingredient for the reduction is the modelling of the command execution $CE$~\cite{LS20, zhu2019, Linnetworked}, which ``transduces" the control commands issued by the supervisor into the events executed by the plant. 
Then, by using the observations $O$, we perform a direct pruning and attack modelling on the synthesized command non-deterministic supervisor $NS$ to obtain $OCNS^{A}$, the supremal safe and observation-consistent command non-deterministic supervisor under sensor-actuator attack. The covertness-breaking states will be encoded in $OCNS^{A}$. In addition, based on the observations $O$, we shall construct $\overline{S^{\downarrow,A}}$, whose marked behavior encodes the least permissive supervisor (that is consistent with $O$) under attack, to make sure that the synthesized sensor-actuator attacker is damage-reachable against any safe supervisor that is consistent with the observations~\cite{LTZS20}.
\item At the second step, based on $G$, $OCNS^{A}$, $\overline{S^{\downarrow,A}}$, $AC$, and $CE^{A}$, we can employ techniques similar to those of~\cite{LS20, LS20J} to reduce the synthesis of maximally permissive covert damage-reachable attackers to the synthesis of maximally permissive safe partial-observation supervisors. 
%\item Optionally, we can perform another step of reduction, which  reduces the synthesis of maximally permissive partial observation supervisor to the normality synthesis in polynomial time, following the same idea as in step 1, if we allow the synthesis of command non-deterministic attackers. 
\end{enumerate}
In particular, the model of the unknown supervisor $S$ is not needed for the above constructions. Intuitively speaking, the synthesized attacker $A$ ensures covertness and damage-reachability because of the following reasons:
\begin{itemize}
\setlength{\itemsep}{3pt}
\setlength{\parsep}{0pt}
\setlength{\parskip}{0pt}
    \item It ensures covertness against all the safe supervisors which are consistent with the observations, since it already ensures covertness against $OCNS^{A}$, the supremal safe and observation-consistent command non-deterministic supervisor (under attack), which embeds all the possible safe (partial-observation) supervisors that are consistent with the observations.
    \item It ensures the damage-reachability against all the supervisors that are consistent with the observations, since it already ensures damage-reachability against $\overline{S^{\downarrow,A}}$~\cite{LTZS20}, which induces the smallest marked behavior. 
\end{itemize}
It follows that we can use the many tools and techniques \cite{Susyna} - \cite{Malik07} that have been developed for the synthesis of (maximally permissive) partial-observation supervisors to synthesize (maximally permissive) covert  damage-reachable attackers assisted with the observations. 

\subsection{Solution methodology}
\label{subsec:Solution methodology}

\noindent \textbf{Step 1: Construction of  $NS$}

Firstly, we shall synthesize $NS$, the supremal safe command non-deterministic supervisor\footnote{We can refer to Fig. 8. Instead of employing a command deterministic supervisor $BT(S)$ (to control $G|| CE$), which issues a unique control  command at each control state, we can employ $NS$ for the control of $G|| CE$. In particular, $NS$ has the choice of issuing different control commands at each control state.}. The procedure is given as follows:

\noindent \textbf{Procedure 1:}
\begin{enumerate}[1.]
\setlength{\itemsep}{3pt}
\setlength{\parsep}{0pt}
\setlength{\parskip}{0pt}
    \item Compute $\mathcal{P} = G||CE = (Q_{\mathcal{P}}, \Sigma_{\mathcal{P}} = \Sigma \cup \Gamma, \xi_{\mathcal{P}}, q_{\mathcal{P}}^{init})$\footnote{By definition, the marker state set of $\mathcal{P}$ should be $Q_{d} \times Q_{ce}$, but here, we shall mark all the states of $\mathcal{P}$ since we only care about safe supervisors.}.
    \item Generate $\mathcal{P}_{r} = (Q_{\mathcal{P}_{r}}, \Sigma_{\mathcal{P}_{r}}, \xi_{\mathcal{P}_{r}}, q_{\mathcal{P}_{r}}^{init})$
    \begin{itemize}
    \setlength{\itemsep}{3pt}
    \setlength{\parsep}{0pt}
    \setlength{\parskip}{0pt}
        \item $Q_{\mathcal{P}_{r}} = Q_{\mathcal{P}} - \{(q, q_{ce}) \in Q_{\mathcal{P}}|\, q \in Q_{d}\}$
        \item $\Sigma_{\mathcal{P}_{r}} = \Sigma_{\mathcal{P}} = \Sigma \cup \Gamma$
        \item $(\forall q, q' \in Q_{\mathcal{P}_{r}})(\forall \sigma \in \Sigma_{\mathcal{P}_{r}}) \, \xi_{\mathcal{P}}(q, \sigma) = q' \Leftrightarrow \xi_{\mathcal{P}_{r}}(q, \sigma) = q'$
        \item $q_{\mathcal{P}_{r}}^{init} = q_{\mathcal{P}}^{init}$
    \end{itemize}
    \item Synthesize the supremal safe supervisor $NS = (Q_{ns}, \Sigma_{ns} = \Sigma \cup \Gamma, \xi_{ns}, q_{ns}^{init})$ over the control constraint $(\Gamma - \{\Sigma_{uc}\}, \Sigma_{o} \cup \Gamma)$ by treating $\mathcal{P}$ as the plant and $\mathcal{P}_{r}$ as the requirement such that $\mathcal{P}||NS$ is safe w.r.t. $\mathcal{P}_{r}$.
\end{enumerate}
We shall briefly explain \textbf{Procedure 1}. As illustrated in Fig. \ref{fig:Supervisory_Control_Bipartite_Supervisor}, at Step 1, we shall construct a lifted plant $\mathcal{P} = G||CE$, where the issuing of different control commands is modelled and can be controlled. In addition, since we only consider safe supervisors, at Step 2, we shall then remove any state of $\{(q, q_{ce}) \in Q_{\mathcal{P}}|\, q \in Q_{d}\}$ in $\mathcal{P}$ to generate the requirement $\mathcal{P}_{r}$. Thus, by treating $\mathcal{P}$ as the plant and $\mathcal{P}_{r}$ as the requirement, we can synthesize the supremal safe command-nondeterministic supervisor $NS$ over the control constraint $(\Gamma - \{\Sigma_{uc}\}, \Sigma_{o} \cup \Gamma)$, whose existence is guaranteed since $\Gamma - \{\Sigma_{uc}\} \subseteq \Sigma_{o} \cup \Gamma$~\cite{WMW10},~\cite{GLM20},~\cite{zhu2019},~\cite{Linnetworked}. Here, the control command $\Sigma_{uc}$ is not controllable to the supervisor because it entirely consists of uncontrollable events, which are always allowed to be fired at the plant $G$.
%It is noteworthy that for any supervisor $S$, the synthesized $NS$ contains the structure $BT(S)^{1}$, whose construction procedure is introduced in Section \ref{subsec:unknown supervisor}.

Based on the structure of $G$ and $CE$, the synthesized $NS$ is a bipartite structure (introduced in Section \ref{subsec:unknown supervisor}). For technical convenience, we shall write the state set of $NS$ as $Q_{ns} = Q_{ns}^{rea} \cup Q_{ns}^{com}$ ($Q_{ns}^{rea}$ and $Q_{ns}^{com}$ denote the set of reaction states and control states, respectively), where 
\begin{itemize}
\setlength{\itemsep}{3pt}
\setlength{\parsep}{0pt}
\setlength{\parskip}{0pt}
    \item At any state of $Q_{ns}^{rea}$, any event in $\Gamma$ is not defined.
    \item At any state of $Q_{ns}^{rea}$, any event in $\Sigma_{uo}$, if defined, leads to self-loops, and any event in $\Sigma_{o}$, if defined, would lead to a transition to a control state.
    \item At any state of $Q_{ns}^{com}$, any event in $\Sigma$ is not defined.
    \item At any state of $Q_{ns}^{com}$, any event in $\Gamma$, if defined, would lead to a transition to a reaction state.
\end{itemize}
Based on \textbf{Procedure 1}, we know that $|Q_{ns}| \leq 2^{|Q| \times |Q_{ce}|}$.

%{\color{red} It may be harder to control the result of the synthesis; but by the definitioin of a supervisor, it seems that at any state of $NS$, whether control state or reaction state, uncontrollable events must be defined and unobservable events if defined lead to self-loops. I see that the discussion is only valid if you compute $G|| CE || NS$, which is the synthesized result in supremica and is a supervisor in other paper's definitions, but we do not refer to this as a supervisor..in this paper}

\begin{figure}[htbp]
\begin{center}
\includegraphics[height=3.3cm]{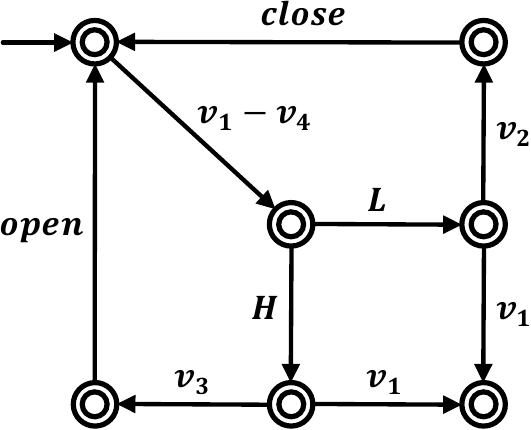}   
\caption{The synthesized supremal safe command-nondeterministic supervisor $NS$}
\label{fig:Example_NS}
\end{center}        
\end{figure}

\textbf{Example IV.1} We shall continue with the water tank example, whose setup is shown in \textbf{Example III.1}. Based on the plant $G$ and command execution automaton $CE$ illustrated in Fig. \ref{fig:Plant G} and Fig. \ref{fig:Example_command execution}, respectively, the synthesized supremal safe command-nondeterministic supervisor $NS$ is illustrated in Fig. \ref{fig:Example_NS}.

\vspace{0.1cm}

\noindent \textbf{Step 2: Construction of $OCNS^{A}$}

Next, based on the synthesized $NS$ and observations $O$, we shall construct $OCNS^{A}$, the supremal safe and observation-consistent command non-deterministic supervisor under sensor-actuator attack. The step-by-step construction procedure is given as follows, including \textbf{Step 2.1} - \textbf{Step 2.3}.

\vspace{0.1cm}

\noindent \textbf{Step 2.1: Construction of $OC$}

\vspace{0.1cm}

Based on the model $M_{o} = (Q_{o}, \Sigma_{o}, \xi_{o}, q_{o}^{init})$ which captures the observations $O$, we shall construct a bipartite structure $OC$ to embed any supervisor consistent with $O$. The construction of $OC$ is similar to that of a bipartite supervisor shown in Section \ref{subsec:unknown supervisor}, which is given as follows:
\[
OC = (Q_{oc}, \Sigma_{oc}, \xi_{oc}, q_{oc}^{init})
\]
\begin{enumerate}[1.]
\setlength{\itemsep}{3pt}
\setlength{\parsep}{0pt}
\setlength{\parskip}{0pt}
    \item $Q_{oc} = Q_{o} \cup Q_{o}^{com} \cup \{q_{oc}^{dump}\}$, where $Q_{o}^{com}:= \{q^{com}|q \in Q_{o}\}$ 
    \item $\Sigma_{oc} = \Sigma \cup \Gamma$
    \item $(\forall q^{com} \in Q_{o}^{com})(\forall \gamma \in \Gamma) \, En_{M_{o}}(q) \subseteq \gamma \Rightarrow \xi_{oc}(q^{com}, \gamma) = q$
    \item $(\forall q \in Q_{o})(\forall \sigma \in \Sigma_{o}) \, \xi_{o}(q, \sigma)! \Rightarrow \xi_{oc}(q, \sigma) = (\xi_{o}(q, \sigma))^{com}$
    \item $(\forall q \in Q_{o})(\forall \sigma \in \Sigma_{o}) \, \neg \xi_{o}(q, \sigma)! \Rightarrow \xi_{oc}(q, \sigma) = q_{oc}^{dump}$
    \item $(\forall q \in Q_{o})(\forall \sigma \in \Sigma_{uo}) \, \xi_{oc}(q, \sigma) = q$
    \item $(\forall \sigma \in \Sigma \cup \Gamma) \, \xi_{oc}(q_{oc}^{dump}, \sigma) = q_{oc}^{dump}$
    \item $q_{oc}^{init} = (q_{o}^{init})^{com}$
\end{enumerate}
Firstly, at Step 1, the state set $Q_{oc}$ consists of the reaction state set $Q_{o}$ and the control state set $Q_{o}^{com}$. In addition, we add a new state $q_{oc}^{dump}$ to denote that some event sequence which is not collected in the observations $O$ happens. At Step 3, for any control state $q^{com}$, we shall allowing the issuing of any control command $\gamma$ satisfying the condition $En_{M_{o}}(q) \subseteq \gamma$, i.e., $\gamma$
can generate the event executions $En_{M_{o}}(q)$ that have been collected at the state $q$. 
%Here, it is noteworthy that the control commands satisfying $En_{M_{o}}(q) \subseteq \gamma$ are only possible to generate the observations $En_{M_{o}}(q)$, but not necessarily to generate such observations because the construction of $OC$ have not taken the structure of $G||CE$ into consideration. 
At Step 4, all the transitions originally defined at the state $q$ in $M_{o}$ are retained and would drive the state change to the control state $(\xi_{o}(q,\sigma))^{com}$. At Step 5, for any reaction state $q$, any event in $\Sigma_{o}$, which has not been collected in the current observations, would lead to a transition to the dump state $q_{oc}^{dump}$ %{\color{red} do you add those observable uncontrollable events that are definitely possible in $G$ to some non-dump state; these observations although are not directly observed, but they can be viewed as being observed and thus enlarge $O$. }. 
%{\color{blue}In $OC$, these events would transit to $q_{oc}^{dump}$. Since they are not collected in $O$, we are not sure what event would happen afterwards, so we need to use $OC$ to do synchronous product with $NS$, then we can obtain the structure containing all possible supervisors consistent with $O$.} {\color{red} I understand the need of completing the observation, but need to check how dump state is treated later..}
At Step 6, at any reaction state, all the events in $\Sigma_{uo}$ will lead to self-loops because they are unobservable. 
At Step 7, any event in $\Sigma \cup \Gamma$ is defined at the state $q_{oc}^{dump}$ since $q_{oc}^{dump}$ is a state denoting that the transition has gone out of the observations collected by the attacker, implying that any event in $\Sigma \cup \Gamma$ might happen.

\vspace{0.1cm}

\noindent \textbf{Step 2.2: Construction of $OCNS$}

\vspace{0.1cm}

We shall adopt the above-constructed $OC$, which embeds any supervisor consistent with $O$, to refine the structure of $NS = (Q_{ns}^{rea} \cup Q_{ns}^{com}, \Sigma_{ns} = \Sigma \cup \Gamma, \xi_{ns}, q_{ns}^{init})$, which encodes all the possible safe bipartite supervisors. To achieve this goal, we compute the synchronous product $OCNS = NS||OC = (Q_{ocns}, \Sigma_{ocns}, \xi_{ocns}, q_{ocns}^{init})$, where it can be easily checked that $Q_{ocns} \subseteq (Q_{ns}^{rea} \times (Q_{o} \cup \{q_{oc}^{dump}\})) \cup (Q_{ns}^{com} \times (Q_{o}^{com} \cup \{q_{oc}^{dump}\}))$.
%{\color{red} is $OCNS$ the supremal safe and observation consistent command non-determinsitic supervisor proven and used later, or simply intuitive explained..(NS is clear, OCNS may require some proof) i need to check later}
\vspace{0.1cm}

\noindent \textbf{Step 2.3: Construction of $OCNS^{A}$}

\vspace{0.1cm}

Based on $OCNS$, we shall encode the effects of the sensor-actuator attacks to generate $OCNS^{A}$, which is similar to the construction procedure of $BT(S)^A$ given in \textbf{Step 2} of Section \ref{subsec:unknown supervisor}.
\[
OCNS^{A} = (Q_{ocns}^{a}, \Sigma_{ocns}^{a}, \xi_{ocns}^{a}, q_{ocns}^{init,a})
\]
\begin{enumerate}[1.]
\setlength{\itemsep}{3pt}
\setlength{\parsep}{0pt}
\setlength{\parskip}{0pt}
    \item $Q_{ocns}^{a} = Q_{ocns} \cup \{q_{cov}^{brk}\}$
    \item $\Sigma_{ocns}^{a} = \Sigma \cup \Sigma_{s,a}^{\#} \cup \Gamma$
    \item $(\forall q, q' \in Q_{ocns}^{a})(\forall \sigma \in \Sigma_{s,a}) \, \xi_{ocns}(q, \sigma) = q' \Rightarrow \xi_{ocns}^{a}(q, \sigma^{\#}) = q' \wedge \xi_{ocns}^{a}(q, \sigma) = q$
    \item $(\forall q \in Q_{ocns}^{a})(\forall \sigma \in \Sigma_{c,a} \cap (\Sigma_{uo} \cup \Sigma_{s,a})) \, q \in Q_{ns}^{rea} \times (Q_{o} \cup \{q_{oc}^{dump}\}) \Rightarrow \xi_{ocns}^{a}(q, \sigma) = q$
    \item $(\forall q, q' \in Q_{ocns}^{a})(\forall \sigma \in (\Sigma - \Sigma_{s,a}) \cup \Gamma) \, \xi_{ocns}(q, \sigma) = q' \Rightarrow \xi_{ocns}^{a}(q, \sigma) = q'$
    \item $(\forall q \in Q_{ocns}^{a})(\forall \sigma \in \Sigma_{o} - \Sigma_{s,a}) \, q \in Q_{ns}^{rea} \times (Q_{o} \cup \{q_{oc}^{dump}\}) \wedge \neg \xi_{ocns}(q, \sigma)! \Rightarrow \xi_{ocns}^{a}(q, \sigma) = q_{cov}^{brk}$
    \item $(\forall q \in Q_{ocns}^{a})(\forall \sigma \in \Sigma_{s,a}) \, q \in Q_{ns}^{rea} \times (Q_{o} \cup \{q_{oc}^{dump}\}) \wedge \neg \xi_{ocns}(q, \sigma)! \Rightarrow \xi_{ocns}^{a}(q, \sigma^{\#}) = q_{cov}^{brk}$
    %\item $(\forall \sigma \in \Sigma \cup \Sigma_{s,a}^{\#} \cup \Gamma) \, \xi_{ocns}^{a}(q_{cov}^{brk}, \sigma) = q_{cov}^{brk}$
    \item $q_{ocns}^{init,a} = q_{ocns}^{init}$
\end{enumerate}
At Step 1, all the states in $OCNS$ are retained, and we shall add a new state $q_{cov}^{brk}$ to denote the covertness-breaking situations.
At Step 3, due to the existence of sensor attack, at any state $q$, any transition labelled by $\sigma \in \Sigma_{s,a}$ in $OCNS$ is relabelled with $\sigma^{\#} \in \Sigma_{s,a}^{\#}$ because the supervisor can observe $\Sigma_{s,a}^{\#}$ instead of $\Sigma_{s,a}$; in addition, a self-loop labelled by $\sigma$ is added at the state $q$ because such event can happen and is unobservable to the supervisor. At Step 4, for any state $q \in Q_{ns}^{rea} \times (Q_{o} \cup \{q_{oc}^{dump}\})$, we shall add the self-loop transitions labelled by events in $\Sigma_{c,a} \cap (\Sigma_{uo} \cup \Sigma_{s,a})$ since they can be enabled due to the actuator attack and are unobservable to the supervisor. At Step 5, all the other transitions, labelled by events in $(\Sigma-\Sigma_{s, A}) \cup \Gamma$, defined in $OCNS$ are kept.
At Step 6 and Step 7, we shall explicitly encode the covertness-breaking situations: at any state $q \in Q_{ns}^{rea} \times (Q_{o} \cup \{q_{oc}^{dump}\})$, any event in $\Sigma_{o}$, which should not have been observed in the absence of attack, denoted by $\neg \xi_{ocns}(q, \sigma)!$, would lead to a transition labelled as $\sigma \in \Sigma_{o} - \Sigma_{s,a}$ or $\sigma^{\#} \in \Sigma_{s,a}^{\#}$ to the state $q_{cov}^{brk}$, meaning that the existence of the sensor-actuator attacker is exposed. 

Based on the model of $OCNS^{A}$, we know that $|Q_{ocns}^{a}| = |Q_{ns}| \times (2|Q_{o}| + 1) + 1 \leq 2^{|Q| \times |Q_{ce}|} \times (2|Q_{o}| + 1) + 1$.

\vspace{0.1cm}

\textbf{Example IV.2} We shall continue with the water tank example. Based on  \textbf{Step 2.1} - \textbf{Step 2.3}, the observation automaton $M_{o}$ and the synthesized supremal safe command-nondeterministic supervisor $NS$ shown in Fig. \ref{fig:Observations M_o} and Fig. \ref{fig:Example_NS}, respectively, we can obtain $OCNS^{A}$, which is illustrated in Fig. \ref{fig:Example_OCNS^{A}}.

\begin{figure}[htbp]
\begin{center}
\includegraphics[height=6cm]{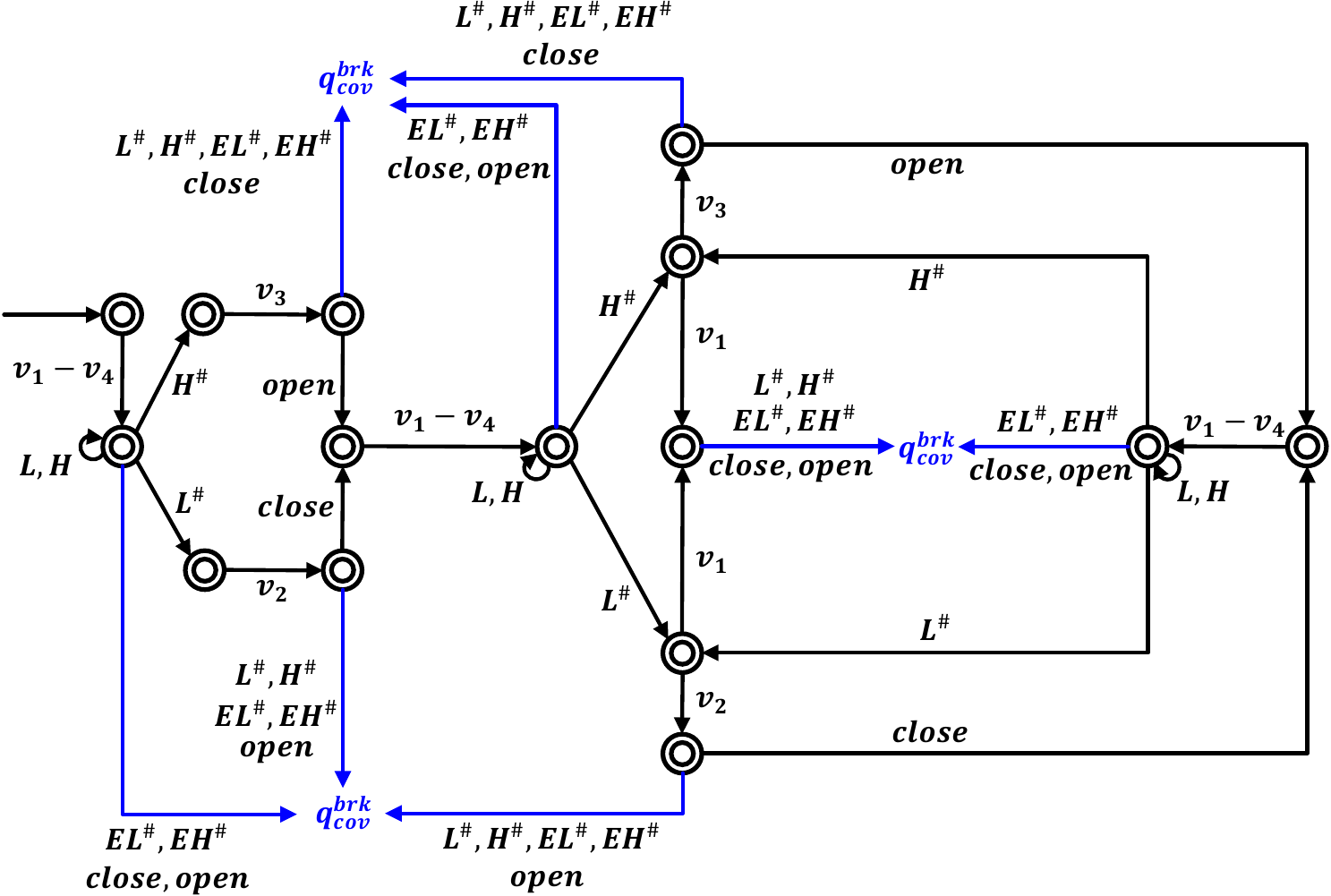}   
\caption{The constructed $OCNS^{A}$}
\label{fig:Example_OCNS^{A}}
\end{center}        
\end{figure}

\vspace{0.1cm}

\emph{Theorem IV.1:} Given a set of observations $O$, for any safe supervisor $S$ that is consistent with $O$, i.e., $O \subseteq P_{o}(L(G||CE||BT(S)))$, it holds that $L(BT(S)^{A}) \subseteq L(OCNS^{A})$. 
%{\color{red} Is there a simpler proof as $BT(S)$ can be embedded in $OCNS$, so their attacked versions, following the same constructions, can be embedded in the same way. This proof that requires versions 1, 2 may complicate the matter} 

\emph{Proof:} Firstly, we prove $L(BT(S)^{1}) \subseteq L(OCNS) = L(NS||OC)$. To prove this result, we only need to show $L(BT(S)) \subseteq L(OCNS) = L(NS||OC)$ as $L(BT(S)^{1}) \subseteq L(BT(S))$. It is straightforward that $L(BT(S)) \subseteq L(NS)$ as $BT(S)$ is safe and $NS$ is the supremal safe command-nondeterministic supervisor. It is also clear that $L(BT(S)) \subseteq L(OC)$, as $BT(S)$ is consistent with observation $O$ and $OC$ embeds any supervisor that is consistent with $O$. Thus, $L(BT(S)^{1}) \subseteq L(BT(S)) \subseteq L(NS) \cap L(OC) = L(OCNS)$.

For any string $s \in L(BT(S)^{1})$ of the form $s_1\gamma$, where $\gamma \in \Gamma$, based on the above analysis, we have $s \in L(OCNS) = L(OC) \cap L(NS)$. 
Thus, we have $En_{OCNS}(\xi_{ocns}(q_{ocns}^{init}, s)) = En_{OC}(\xi_{oc}(q_{oc}^{init}, s)) \cap En_{NS}(\xi_{ns}(q_{ns}^{init}, s)) = En_{NS}(\xi_{ns}(q_{ns}^{init}, s))$ as any event in $\Sigma$ is defined at the state $\xi_{oc}(q_{oc}^{init}, s)$ of $OC$ by construction. Since $NS \Vert P_{\Sigma_o \cup \Gamma}(G\lVert CE) = NS$ and $BT(S)^1 = BT(S) \lVert P_{\Sigma_o \cup \Gamma}(G\lVert CE)$, we have $En_{OCNS}(\xi_{ocns}(q_{ocns}^{init}, s)) = En_{NS}(\xi_{ns}(q_{ns}^{init}, s)) = En_{BT(S)^{1}}(\xi_{bs,1}(q_{bs,1}^{init}, s))$. Since \textbf{Step 2} of constructing $BT(S)^{A}$ based on $BT(S)^{1}$ in Section \ref{subsec:unknown supervisor} and \textbf{Step 2.3} of constructing $OCNS^{A}$ based on $OCNS$ follow the same procedures, we conclude that $L(BT(S)^{A}) \subseteq L(OCNS^{A})$. This completes the proof. \hfill $\blacksquare$

\vspace{0.1cm}

\emph{Corollary IV.1:} For any safe supervisor $S$ consistent with $O$, if $s \in L(BT(S)^{A})$, it holds that $\xi_{bs,a}(q_{bs,a}^{init}, s) = q^{detect} \Leftrightarrow \xi_{ocns}^{a}(q_{ocns}^{init,a}, s) = q_{cov}^{brk}$. 

\emph{Proof:} This follows from the analysis given in the proof of \emph{Theorem IV.1}. \hfill $\blacksquare$

\vspace{0.1cm}

\noindent \textbf{Step 3: Construction of $\overline{S^{\downarrow,A}}$}

\vspace{0.1cm}

Next, we shall construct $\overline{S^{\downarrow,A}}$~\cite{LTZS20}, a complete automaton whose marked behavior models the least permissive supervisor (that is consistent with $O$) under attack. The step-by-step construction procedure is given as follows, including \textbf{Step 3.1} - \textbf{Step 3.3}.

\vspace{0.1cm}

\noindent \textbf{Step 3.1: Construction of $S^{\downarrow}$}

\vspace{0.1cm}

Based on the model $M_{o} = (Q_{o}, \Sigma_{o}, \xi_{o}, q_{o}^{init})$ that captures the observations $O$, we shall construct the least permissive supervisor $S^{\downarrow}$~\cite{LTZS20} which is consistent with $O$. Let
\[
S^{\downarrow} = (Q_{s}^{\downarrow}, \Sigma_{s}^{\downarrow}, \xi_{s}^{\downarrow}, q_{s}^{init,\downarrow})
\]
\begin{enumerate}[1.]
\setlength{\itemsep}{3pt}
\setlength{\parsep}{0pt}
\setlength{\parskip}{0pt}
    \item $Q_{s}^{\downarrow} = Q_{o}$
    \item $\Sigma_{s}^{\downarrow} = \Sigma$
    \item $(\forall q, q' \in Q_{o})(\forall \sigma \in \Sigma_{o}) \, \xi_{o}(q, \sigma) = q' \Rightarrow \xi_{s}^{\downarrow}(q, \sigma) = q'$
    \item $(\forall q \in Q_{o})(\forall \sigma \in \Sigma_{uc} \cap \Sigma_{uo} = \Sigma_{uo}) \, \xi_{s}^{\downarrow}(q, \sigma) = q$ 
    \item $(\forall q \in Q_{o})(\forall \sigma \in \Sigma_{uc} \cap \Sigma_{o}) \, \neg \xi_{o}(q, \sigma)! \Rightarrow \xi_{s}^{\downarrow}(q, \sigma) = q_{o}^{dl}$
    \item $q_{s}^{init,\downarrow} = q_{o}^{init}$
\end{enumerate}
At Step 3, we shall retain all the transitions originally defined in $M_{o}$. Then, at any state $q \in Q_{o}$, Step 4 and Step 5 would complete the undefined transitions labelled by events in $\Sigma_{uc}$ to satisfy the controllability, where the unobservable parts would lead to self-loops at Step 4 to satisfy the observability, and the observable parts would transit to the deadlocked state $q_{o}^{dl}$ at Step 5. 

\emph{Theorem IV.2:} Given a set of observations $O$, $S^{\downarrow}$ is the least permissive supervisor among all the supervisors that are consistent with $O$.  %{\color{red} instead of doing the proof here, can cite ~\cite{LTZS20} directly to remove the proof.}

\emph{Proof:} Firstly, we prove $S^{\downarrow}$ is consistent with $O$. Since $O$ is a finite set of observations of the executions of $G||S$, we have $O \subseteq P_o(L(G|| S))$. Thus, $O \subseteq P_o(L(G))$. Based on the fact that $L(S^{\downarrow}) = P_{o}^{-1}(O(\Sigma_{uc} \cap \Sigma_o)^*)$, we have $P_o(L(G|| S^{\downarrow})) = P_o(L(G) \cap P_{o}^{-1}(O(\Sigma_{uc} \cap \Sigma_o)^*)) = P_o(L(G)) \cap O(\Sigma_{uc} \cap \Sigma_o)^*\supseteq O$.

Secondly, we prove $S^{\downarrow}$ is the least permissive supervisor that is consistent with $O$. We use the fact that every supervisor $S$ over the control constraint $(\Sigma_c, \Sigma_o)$, where $\Sigma_c \subseteq \Sigma_o$, satisfies $L(S)=P_o^{-1}(P_o(L(S))(\Sigma_{uc} \cap \Sigma_{o})^*)~$\cite{Lin15}.
Thus, for any supervisor $S$ that is consistent with $O$, since $O \subseteq P_{o}(L(G||S))= P_{o}(L(G) \cap P_{o}^{-1}(P_{o}(L(S))(\Sigma_{uc} \cap \Sigma_{o})^{*}))= P_{o}(L(G)) \cap P_{o}(L(S))(\Sigma_{uc} \cap \Sigma_{o})^{*}$, we have $O \subseteq P_{o}(L(S))(\Sigma_{uc} \cap \Sigma_{o})^{*}$. Thus, $L(S^{\downarrow}) = P_{o}^{-1}(O(\Sigma_{uc} \cap \Sigma_{o})^{*}) \subseteq P_{o}^{-1}(P_{o}(L(S))(\Sigma_{uc} \cap \Sigma_{o})^{*}) = L(S)$. This completes the proof. \hfill $\blacksquare$

\vspace{0.1cm}

\noindent \textbf{Step 3.2: Construction of $S^{\downarrow,A}$}

\vspace{0.1cm}
Based on $S^{\downarrow}$, we shall construct $S^{\downarrow,A}$, whose behavior encodes the least permissive supervisor (consistent with $O$) under the effects of the sensor-actuator attack.
\[
S^{\downarrow,A} = (Q_{s}^{\downarrow,a}, \Sigma_{s}^{\downarrow,a}, \xi_{s}^{\downarrow,a}, q_{s}^{init,\downarrow,a})
\]
\begin{enumerate}[1.]
\setlength{\itemsep}{3pt}
\setlength{\parsep}{0pt}
\setlength{\parskip}{0pt}
    \item $Q_{s}^{\downarrow,a} = Q_{s}^{\downarrow} \cup \{q^{risk}\}$
    \item $\Sigma_{s}^{\downarrow,a} = \Sigma \cup \Sigma_{s,a}^{\#}$
    \item $(\forall q, q' \in Q_{s}^{\downarrow,a})(\forall \sigma \in \Sigma_{s,a}) \, \xi_{s}^{\downarrow}(q, \sigma) = q' \Rightarrow \xi_{s}^{\downarrow,a}(q, \sigma^{\#}) = q' \wedge \xi_{s}^{\downarrow,a}(q, \sigma) = q$
    \item $(\forall q \in Q_{s}^{\downarrow,a})(\forall \sigma \in \Sigma_{c,a} \cap (\Sigma_{uo} \cup \Sigma_{s,a})) \, \neg \xi_{s}^{\downarrow}(q, \sigma)! \Rightarrow \xi_{s}^{\downarrow,a}(q, \sigma) = q$
    \item $(\forall q, q' \in Q_{s}^{\downarrow,a})(\forall \sigma \in \Sigma - \Sigma_{s,a}) \, \xi_{s}^{\downarrow}(q, \sigma) = q' \Rightarrow \xi_{s}^{\downarrow,a}(q, \sigma) = q'$
    \item $(\forall q \in Q_{s}^{\downarrow,a})(\forall \sigma \in \Sigma_{o} - \Sigma_{s,a}) \, \neg \xi_{s}^{\downarrow}(q, \sigma)! \Rightarrow \xi_{s}^{\downarrow,a}(q, \sigma) = q^{risk}$
    \item $(\forall q \in Q_{s}^{\downarrow,a})(\forall \sigma \in \Sigma_{s,a}) \, \neg \xi_{s}^{\downarrow}(q, \sigma)! \Rightarrow \xi_{s}^{\downarrow,a}(q, \sigma^{\#}) = q^{risk}$
    \item $q_{s}^{init,\downarrow,a} = q_{s}^{init,\downarrow}$
\end{enumerate} 
The construction of $S^{\downarrow,A}$ is similar to that of $BT(S)^{A}$ in \textbf{Step 2} of Section \ref{subsec:unknown supervisor}, where $q^{risk}$ in $S^{\downarrow,A}$ serves as the same role as $q^{detect}$ in $BT(S)^{A}$.

\vspace{0.1cm}

\noindent \textbf{Step 3.3: Construction of $\overline{S^{\downarrow,A}}$}

\vspace{0.1cm}
Based on $S^{\downarrow,A}$, we shall construct $\overline{S^{\downarrow,A}}$ by performing the completion to make $\overline{S^{\downarrow,A}}$ become a complete automaton, where now only the marked behavior encodes the least permissive supervisor (consistent with $O$) under attack.
\[
\overline{S^{\downarrow,A}} = (\overline{Q_{s}^{\downarrow,a}}, \overline{\Sigma_{s}^{\downarrow,a}}, \overline{\xi_{s}^{\downarrow,a}}, \overline{q_{s}^{init,\downarrow,a}}, \overline{Q_{s,m}^{\downarrow,a}})
\]
\begin{enumerate}[1.]
\setlength{\itemsep}{3pt}
\setlength{\parsep}{0pt}
\setlength{\parskip}{0pt}
    \item $\overline{Q_{s}^{\downarrow,a}} = Q_{s}^{\downarrow,a} \cup \{q^{dump}\}$
    \item $\overline{\Sigma_{s}^{\downarrow,a}} = \Sigma \cup \Sigma_{s,a}^{\#}$
    \item $(\forall q, q' \in \overline{Q_{s}^{\downarrow,a}})(\forall \sigma \in \Sigma \cup \Sigma_{s,a}^{\#}) \, \xi_{s}^{\downarrow,a}(q, \sigma) = q' \Rightarrow \overline{\xi_{s}^{\downarrow,a}}(q, \sigma) = q'$
    \item $(\forall q \in \overline{Q_{s}^{\downarrow,a}})(\forall \sigma \in \Sigma \cup \Sigma_{s,a}^{\#}) \, \neg \xi_{s}^{\downarrow,a}(q, \sigma)! \Rightarrow \overline{\xi_{s}^{\downarrow,a}}(q, \sigma) = q^{dump}$
    \item $(\forall \sigma \in \Sigma \cup \Sigma_{s,a}^{\#}) \, \overline{\xi_{s}^{\downarrow,a}}(q^{dump}, \sigma) = q^{dump}$
    \item $\overline{q_{s}^{init,\downarrow,a}} = q_{s}^{init,\downarrow,a}$
    \item $\overline{Q_{s,m}^{\downarrow,a}} = Q_{s}^{\downarrow,a}$
\end{enumerate}
%At Step 1, we add a new state $q^{dump}$ to denote the deviation from the least permissive supervisor information under attack.
%At Step 3, all the transitions in $S^{\downarrow,a}$ are retained. Then, at Step 4 and Step 5, we shall complete all the undefined transitions labelled by events in $\Sigma \cup \Sigma_{s,a}^{\#}$ to make it become a complete automaton. It is noteworthy that these newly added transitions would transit to a non-marker state $q^{dump}$ because such transitions have jumped out of the least permissive supervisor information that could be used by the attacker. 
Intuitively speaking, as long as the attacker makes use of the marked behavior of $\overline{S^{\downarrow,A}}$ to implement attacks, it can ensure damage-infliction against any (unknown) safe supervisor that is consistent with the observations. Based on the model of $\overline{S^{\downarrow,A}}$, we know that $|\overline{Q_{s}^{\downarrow,a}}| = |Q_{o}| + 2$.

\vspace{0.1cm}

\begin{figure}[htbp]
\begin{center}
\includegraphics[height=5.3cm]{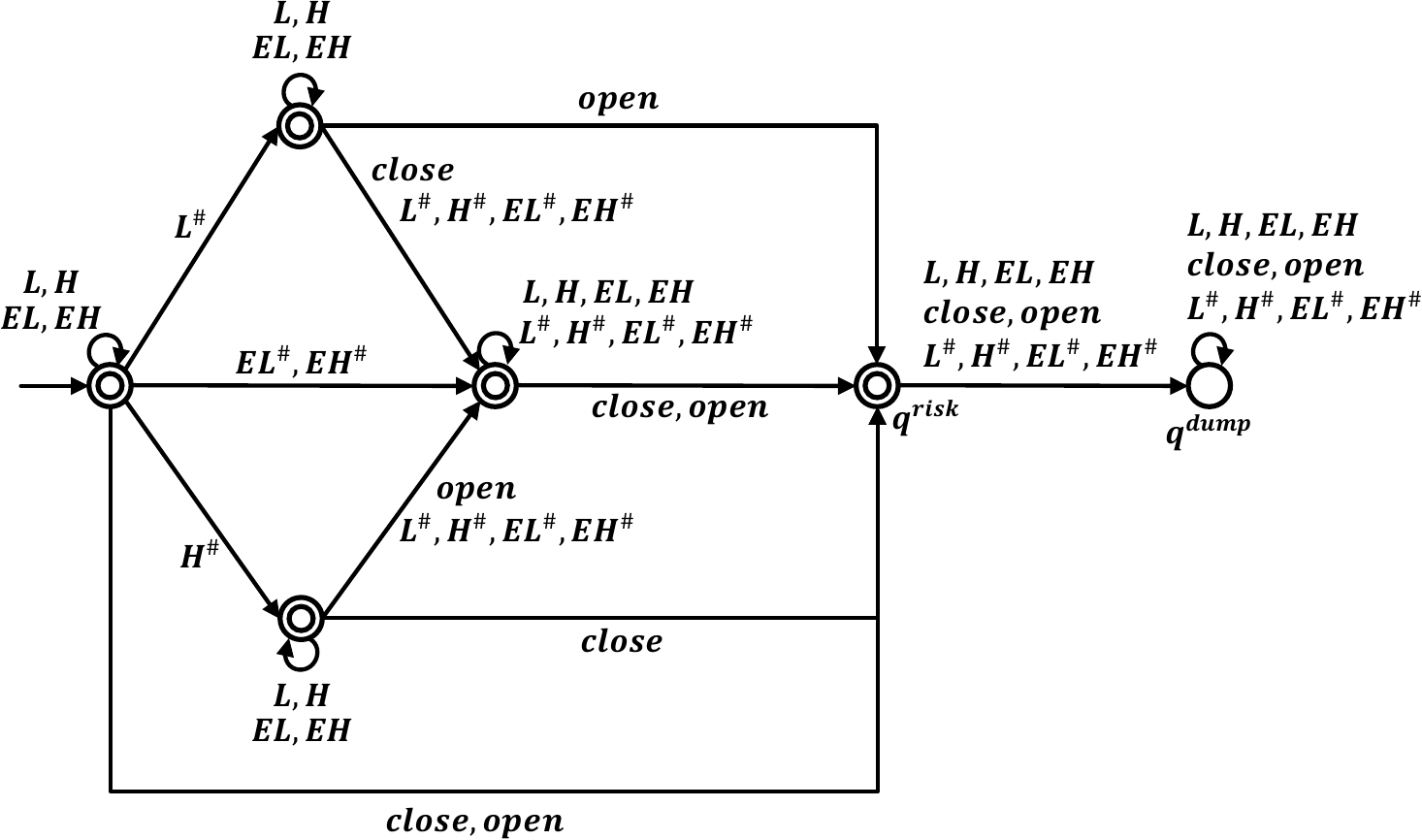}   
\caption{The constructed $\overline{S^{\downarrow,A}}$}
\label{fig:Example_S_downarrow_A}
\end{center}        
\end{figure}

\textbf{Example IV.3} We shall continue with the water tank example. Based on \textbf{Step 3.1} - \textbf{Step 3.3} and the observation automaton $M_{o}$ shown in Fig. \ref{fig:Observations M_o}, we generate $\overline{S^{\downarrow,A}}$, which is illustrated in Fig. \ref{fig:Example_S_downarrow_A}.

\vspace{0.1cm}

\noindent \textbf{Step 4: Synthesis of the sensor-actuator attacker $A$}

\vspace{0.1cm}

Now, we are ready to provide the procedure for the synthesis of maximally permissive covert  damage-reachable sensor-actuator attackers against all the safe supervisors that are consistent with the observations, which is given as follows:

\noindent \textbf{Procedure 2:}
\begin{enumerate}[1.]
\setlength{\itemsep}{3pt}
\setlength{\parsep}{0pt}
\setlength{\parskip}{0pt}
    \item Compute $\mathcal{P} = G||CE^{A}||AC||OCNS^{A}||\overline{S^{\downarrow,A}} = (Q_{\mathcal{P}}, \Sigma_{\mathcal{P}}, \xi_{\mathcal{P}}, q_{\mathcal{P}}^{init}, Q_{\mathcal{P},m})$. 
    \item Generate $\mathcal{P}_{r} = (Q_{\mathcal{P}_{r}}, \Sigma_{\mathcal{P}_{r}}, \xi_{\mathcal{P}_{r}}, q_{\mathcal{P}_{r}}^{init}, Q_{\mathcal{P}_{r},m})$.
    \begin{itemize}
    \setlength{\itemsep}{3pt}
    \setlength{\parsep}{0pt}
    \setlength{\parskip}{0pt}
        \item $Q_{\mathcal{P}_{r}} = Q_{\mathcal{P}} - Q_{1}$
        \begin{itemize}
        \setlength{\itemsep}{3pt}
        \setlength{\parsep}{0pt}
        \setlength{\parskip}{0pt}
            \item $Q_{1} = \{(q, q_{ce,a}, q_{ac}, q_{ocns}^{a}, \overline{q_{s}^{\downarrow,a}}) \in Q_{\mathcal{P}}|\, q \notin Q_{d} \wedge q_{ocns}^{a} = q_{cov}^{brk}\}$
        \end{itemize}
        \item $\Sigma_{\mathcal{P}_{r}} = \Sigma_{\mathcal{P}}$
        \item $(\forall q, q' \in Q_{\mathcal{P}_{r}})(\forall \sigma \in \Sigma_{\mathcal{P}_{r}}) \, \xi_{\mathcal{P}}(q, \sigma) = q' \Leftrightarrow \xi_{\mathcal{P}_{r}}(q, \sigma) = q'$
        \item $q_{\mathcal{P}_{r}}^{init} = q_{\mathcal{P}}^{init}$
        \item $Q_{\mathcal{P}_{r},m} = Q_{\mathcal{P},m} - Q_{1}$
    \end{itemize}
    \item Synthesize a (maximally permissive) supervisor $A = (Q_{a}, \Sigma_{a}, \xi_{a}, q_{a}^{init}, Q_{a,m})$ over the attacker's control constraint $\mathscr{C}_{ac}$ by treating $\mathcal{P}$ as the plant and $\mathcal{P}_{r}$ as the requirement such that $\mathcal{P}||A$ is marker-reachable and safe w.r.t. $\mathcal{P}_{r}$.
\end{enumerate}

We shall briefly explain \textbf{Procedure 2}. At Step 1, we generate a new plant $\mathcal{P} = G||CE^{A}||AC||OCNS^{A}||\overline{S^{\downarrow,A}}$. At Step 2, we generate the requirement $\mathcal{P}_{r}$ from $\mathcal{P}$ by removing those states where the covertness is broken, denoted by $q \notin Q_{d} \wedge q_{ocns}^{a} = q_{cov}^{brk}$. Then we synthesize a (maximally permissive) sensor-actuator attacker at Step 3. Intuitively speaking, 1) since $OCNS^{A}$ have encoded all the safe bipartite supervisors that are consistent with the observations $O$, removing those covertness-breaking states in the requirement $\mathcal{P}_{r}$ can enforce the covertness against any unknown safe supervisor that is consistent with $O$, and 2) since the marked behavior of $\overline{S^{\downarrow,A}}$ encodes the least permissive supervisor under attack, ensuring the marker-reachability for $\mathcal{P}||A$ can enforce that the attacker can always cause damage-infliction against any (unknown) safe supervisor that is consistent with $O$. 

Next, we shall formally prove the correctness of the proposed solution methodology.

\vspace{0.1cm}

\emph{Theorem IV.3:} Given a set of observations $O$, the sensor-actuator attacker $A$ generated in \textbf{Procedure 2}, if non-empty, is covert for any safe supervisor that is consistent with $O$.

\emph{Proof:} We need to prove that, for any safe supervisor $S$, any state in $\{(q, q_{ce,a}, q_{ac}, q_{bs,a}, q_{a}) \in Q \times Q_{ce,a} \times Q_{ac} \times Q_{bs,a} \times Q_{a}|\, q \notin Q_{d} \wedge q_{bs,a} = q^{detect}\}$ is not reachable in $G||CE^{A}||AC||BT(S)^{A}||A$. 

We adopt the contradiction. Suppose some above-mentioned state, where $q \in Q - Q_{d}$, $q_{bs,a} = q^{detect}$, can be reached in $G||CE^{A}||AC||BT(S)^{A}||A$ via some string $s \in L(G||CE^{A}||AC||BT(S)^{A}||A)$. Then, $s$ can be executed in $G$, $CE^{A}$, $AC$, $BT(S)^{A}$, and $A$, after we lift their alphabets to $\Sigma \cup \Sigma_{s,a}^{\#} \cup \Gamma \cup \{stop\}$. Then, based on \emph{Theorem IV.1} and the construction of $\overline{S^{\downarrow,A}}$, which is a complete automaton, we know that $s$ can be executed in $OCNS^{A}$, and $\overline{S^{\downarrow,A}}$. Thus, $s$ can also be executed in $G||CE^{A}||AC||OCNS^{A}||\overline{S^{\downarrow,A}}||A$. Next, we shall check what state is reached in $G||CE^{A}||AC||OCNS^{A}||\overline{S^{\downarrow,A}}||A$ via the string $s$. Clearly, state $q \in Q - Q_{d}$ is reached in $G$ and state $q_{cov}^{brk}$ is reached in $OCNS^{A}$ according to \emph{Corollary IV.1}. Thus, the state $(q, q_{ce,a}, q_{ac}, q_{cov}^{brk}, \overline{q_{s}^{\downarrow,a}}, q_{a})$, where $q \in Q - Q_{d}$, is reached in $G||CE^{A}||AC||OCNS^{A}||\overline{S^{\downarrow,A}}||A$ via the string $s$, which is a contradiction to the fact that $A$ is a safe supervisor for the plant $G||CE^{A}||AC||OCNS^{A}||\overline{S^{\downarrow,A}}$ based on Step 3 of \textbf{Procedure 2}. Then, the supposition does not hold and the proof is completed. \hfill $\blacksquare$

\vspace{0.1cm}

\emph{Theorem IV.4:} Given a set of observations $O$, the sensor-actuator attacker $A$ generated in \textbf{Procedure 2}, if non-empty, is damage-reachable for any safe supervisor that is consistent with $O$.

\emph{Proof:} Firstly, the sensor-actuator attacker $A$ generated in \textbf{Procedure 2}, if non-empty, must satisfy that $\mathcal{P}||A$ is marker-reachable, i.e., some state $(q, q_{ce,a}, q_{ac}, q_{ocns}^{a}, \overline{q_{s}^{\downarrow,a}}, q_{a}) \in Q_{d} \times Q_{ce,a} \times Q_{ac} \times Q_{ocns}^{a} \times \overline{Q_{s,m}^{\downarrow,a}} \times Q_{a}$ can be reached in $G||CE^{A}||AC||OCNS^{A}||\overline{S^{\downarrow,A}}||A$ via some string $s \in L(G||CE^{A}||AC||OCNS^{A}||\overline{S^{\downarrow,A}}||A)$ such that $P_{\Sigma \cup \Sigma_{s,a}^{\#}}(s) \in L_{m}(\overline{S^{\downarrow,A}})$, where $P_{\Sigma \cup \Sigma_{s,a}^{\#}}: (\Sigma \cup \Sigma_{s,a}^{\#} \cup \Gamma \cup \{stop\})^{*} \rightarrow (\Sigma \cup \Sigma_{s,a}^{\#})^{*}$. According to \emph{Theorem IV.2} that $S^{\downarrow}$ is the least permissive supervisor that is consistent with $O$, we know that for any other supervisor $S$ that is consistent with $O$, there always exists a string $s' \in (\Sigma \cup \Sigma_{s,a}^{\#} \cup \Gamma \cup \{stop\})^{*}$ such that $P_{\Sigma \cup \Sigma_{s,a}^{\#}}(s') = P_{\Sigma \cup \Sigma_{s,a}^{\#}}(s)$ and $s'$ can be executed in $G$, $CE^{A}$, $AC$, $BT(S)^{A}$, and $A$ after we lift their alphabets to $\Sigma \cup \Sigma_{s,a}^{\#} \cup \Gamma \cup \{stop\}$. Thus, $s'$ can be executed in $G||CE^{A}||AC||BT(S)^{A}||A$ and the state $q \in Q_{d}$ is reached in $G$ via the string $s'$, which means that some marker state is reachable in $G||CE^{A}||AC||BT(S)^{A}||A$. This completes the proof. \hfill $\blacksquare$

Now, we are ready to show that \textbf{Problem 1}, the main problem to be solved in this work, can be reduced to a Ramadge-Wonham supervisory control problem.%, i.e., the solutions of these two problems are equivalent.

\vspace{0.1cm}

\emph{Theorem IV.5:} Given the plant $G$ and a set of observations $O$, there exists a covert damage-reachable sensor-actuator attacker $A = (Q_{a}, \Sigma_{a}, \xi_{a}, q_{a}^{init})$ w.r.t. the attack constraint $(\Sigma_{o}, \Sigma_{s,a}, \Sigma_{c,a})$ against any safe supervisor that is consistent with $O$ if and only if there exists a supervisor $S'$ over the attacker's control constraint $\mathscr{C}_{ac}$ for the plant $\mathcal{P} = G||CE^{A}||AC||OCNS^{A}||\overline{S^{\downarrow,A}}$ such that 
\begin{enumerate}[a)]
\setlength{\itemsep}{3pt}
\setlength{\parsep}{0pt}
\setlength{\parskip}{0pt}
    \item Any state in $\{(q, q_{ce,a}, q_{ac}, q_{ocns}^{a}, \overline{q_{s}^{\downarrow,a}}, q_{s}') \in Q \times Q_{ce,a} \times Q_{ac} \times Q_{ocns}^{a} \times \overline{Q_{s}^{\downarrow,a}} \times Q_{s}'| \, q \notin Q_{d} \wedge q_{ocns}^{a} = q_{cov}^{brk}\}$ is not reachable in $\mathcal{P}||S'$, where $Q_{s}'$ is the state set of $S'$.
    \item $\mathcal{P}||S'$ is marker-reachable.
\end{enumerate}

\emph{Proof:}  (If) Suppose there exists a supervisor $S'$ over the attacker's control constraint $\mathscr{C}_{ac}$ for the plant $\mathcal{P} = G||CE^{A}||AC||OCNS^{A}||\overline{S^{\downarrow,A}}$ such that the above Condition a) and Condition b) are satisfied. Then, based on \emph{Theorem IV.3} and \emph{Theorem IV.4}, we know that $A=S'$ is a covert  damage-reachable sensor-actuator attacker w.r.t. the attack constraint $(\Sigma_{o}, \Sigma_{s,a}, \Sigma_{c,a})$ against any safe supervisor that is consistent with $O$. This completes the proof of sufficiency.

(Only if) We need to prove that $A$ can satisfy the Condition a) and Condition b) w.r.t. the plant $\mathcal{P}$ and thus we can choose $S'=A$. Firstly, we shall show that any state in $\{(q, q_{ce,a}, q_{ac}, q_{ocns}^{a}, \overline{q_{s}^{\downarrow,a}}, q_{a}) \in Q \times Q_{ce,a} \times Q_{ac} \times Q_{ocns}^{a} \times \overline{Q_{s}^{\downarrow,a}} \times Q_{a}| \, q \notin Q_{d} \wedge q_{ocns}^{a} = q_{cov}^{brk}\}$ is not reachable in $\mathcal{P}||A = G||CE^{A}||AC||OCNS^{A}||\overline{S^{\downarrow,A}}||A$. We carry out the proof by  contradiction and suppose that some state $(q, q_{ce,a}, q_{ac}, q_{ocns}^{a}, \overline{q_{s}^{\downarrow,a}}, q_{a}) \in Q \times Q_{ce,a} \times Q_{ac} \times Q_{ocns}^{a} \times \overline{Q_{s}^{\downarrow,a}} \times Q_{a}$, where $q \in Q - Q_{d}$, $q_{ocns}^{a} = q_{cov}^{brk}$ can be reached in $G||CE^{A}||AC||OCNS^{A}||\overline{S^{\downarrow,A}}||A$ via a string $s$. Thus, $s$ can be executed in $G$, $CE^{A}$, $AC$, $OCNS^{A}$, and $A$ after we lift their alphabets to $\Sigma \cup \Sigma_{s,a}^{\#} \cup \Gamma \cup \{stop\}$. Since $OCNS^{A}$ only embeds all the safe bipartite supervisors that are consistent with $O$ (under attack), we can always find a safe supervisor $S$ that is consistent with $O$ such that $s$ can be executed in $BT(S)^{A}$ and the state $q^{detect}$ is reached in $BT(S)^{A}$ via the string $s$. Thus, in $G||CE^{A}||AC||BT(S)^{A}||A$, the state $(q, q_{ce,a}, q_{ac}, q_{bs,a}, q_{a}) \in Q \times Q_{ce,a} \times Q_{ac} \times Q_{bs,a} \times Q_{a}$, where $q \in Q - Q_{d}$, $q_{bs,a} = q^{detect}$, can be reached via the string $s$, and this causes the contradiction with the fact that $A$ is covert against against any safe supervisor that is consistent with $O$. Thus, the supposition does not hold.

Secondly, since $A$ is damage-reachable against any safe supervisor that is consistent with $O$, we know that $A$ is also damage-reachable against $S^{\downarrow}$, the least permissive supervisor that is consistent with $O$ based on \emph{Theorem IV.2}. Thus, $G||CE^{A}||AC||BT(S^{\downarrow})^{A}||A$ is marker-reachable, i.e., some state $(q, q_{ce,a}, q_{ac}, q_{bs,a}, q_{a}) \in Q_{d} \times Q_{ce,a} \times Q_{ac} \times Q_{bs,a} \times Q_{a}$ can be reached in $G||CE^{A}||AC||BT(S^{\downarrow})^{A}||A$ via some string $s \in L(G||CE^{A}||AC||BT(S^{\downarrow})^{A}||A)$. Then, $s$ can be executed in $G$, $CE^{A}$, $AC$, $BT(S^{\downarrow})^{A}$, and $A$, after we lift their alphabets to $\Sigma \cup \Sigma_{s,a}^{\#} \cup \Gamma \cup \{stop\}$. Based on \emph{Theorem IV.1} and the construction of $\overline{S^{\downarrow,A}}$, which is a complete automaton, we know that $s$ can be executed in $OCNS^{A}$ and $\overline{S^{\downarrow,A}}$, and $P_{\Sigma \cup \Sigma_{s,a}^{\#}}(s) \in L_{m}(\overline{S^{\downarrow,A}})$. Thus, some state $(q, q_{ce,a}, q_{ac}, q_{ocns}^{a}, \overline{q_{s}^{\downarrow,a}}, q_{a}) \in Q_{d} \times Q_{ce,a} \times Q_{ac} \times Q_{ocns}^{a} \times \overline{Q_{s,m}^{\downarrow,a}} \times Q_{a}$ is reached in $G||CE^{A}||AC||OCNS^{A}||\overline{S^{\downarrow,A}}||A$ via the string $s$, i.e., $G||CE^{A}||AC||OCNS^{A}||\overline{S^{\downarrow,A}}||A$ is marker-reachable. This completes the proof of necessity. \hfill $\blacksquare$

\vspace{0.1cm}

\emph{Theorem IV.6:} The sensor-actuator attacker $A$ generated in \textbf{Procedure 2}, if non-empty, is a solution for \textbf{Problem 1}.

\emph{Proof:}  Based on \emph{Theorem IV.5}, the problem of synthesizing a covert  damage-reachable sensor-actuator attacker $A$ w.r.t. the attack constraint $(\Sigma_{o}, \Sigma_{s,a}, \Sigma_{c,a})$ against any safe  supervisor that is consistent with the observations $O$ has been reduced to a Ramadge-Wonham supervisory control problem formulated at Step 3 of \textbf{Procedure 2}. Thus, \textbf{Procedure 2} can synthesize a (maximally permissive) covert damage-reachable sensor-actuator attacker against any (unknown) safe supervisor that is consistent with the observations $O$ for \textbf{Problem 1}. \hfill $\blacksquare$

\vspace{0.1cm}

\emph{Theorem IV.7:} The supremal covert  damage-reachable sensor-actuator attacker against any safe supervisor that is consistent with $O$ exists.

\emph{Proof:} This is straightforward based on the attacker's control constraint $\mathscr{C}_{ac} = (\Sigma_{c,a} \cup \Sigma_{s,a}^{\#} \cup \{stop\}, \Sigma_{o} \cup \Sigma_{s,a}^{\#} \cup \{stop\})$ and the fact that $\Sigma_{c, a} \subseteq \Sigma_c \subseteq \Sigma_o$. \hfill $\blacksquare$

\vspace{0.1cm}

\emph{Theorem IV.8:} \textbf{Problem 1} is decidable.

\emph{Proof:}  Based on \emph{Theorem IV.5} and \emph{Theorem IV.6}, and the fact that \textbf{Procedure 2} terminates within finite steps, we immediately have this result. \hfill $\blacksquare$

\vspace{0.1cm}

Next, we shall analyze the computational complexity of the proposed algorithm to synthesize a covert  damage-reachable sensor-actuator attacker, which depends on the complexity of two synthesis steps: \textbf{Procedure 1} and \textbf{Procedure 2}. By using the normality based synthesis approach \cite{WMW10,WLLW18}, the complexity of \textbf{Procedure 1} and \textbf{Procedure 2} are $O((|\Sigma|+|\Gamma|)2^{|Q|\times|Q_{ce}|})$ and $O((|\Sigma| + |\Gamma|)2^{|Q_{\mathcal{P}}|})$, respectively, where
\begin{itemize}
\setlength{\itemsep}{3pt}
\setlength{\parsep}{0pt}
\setlength{\parskip}{0pt}
    %\item %$|\Sigma_{\mathcal{P}}| = |\Sigma| + |\Sigma_{s,a}| + |\Gamma| + 1$
    %$|\Sigma_{\mathcal{P}}| = |\Sigma| + |\Gamma|$
    \item $|Q_{\mathcal{P}}| = |Q| \times |Q_{ce,a}| \times |Q_{ac}| \times |Q_{ocns}^{a}| \times |\overline{Q_{s}^{\downarrow,a}}|$
    \item $|Q_{ce}| = |Q_{ce,a}| = |\Gamma| + 1$
    \item $|Q_{ac}| = 3$
    \item $|Q_{ocns}^{a}| \leq 2^{|Q| \times |Q_{ce}|} \times (2|Q_{o}| + 1) + 1$
    \item $|\overline{Q_{s}^{\downarrow,a}}| = |Q_{o}| + 2$
\end{itemize}
Thus, the computational complexity of the proposed algorithm is $O((|\Sigma|+|\Gamma|)2^{|Q|\times|Q_{ce}|} + (|\Sigma| + |\Gamma|)2^{|Q_{\mathcal{P}}|}) = O((|\Sigma| + |\Gamma|)2^{|Q_{\mathcal{P}}|})$.

\emph{Remark IV.1:} If we remove the assumption that $\Sigma_c \subseteq \Sigma_o$, then \textbf{Procedure 2} is still a sound procedure for \textbf{Problem 1} but it is in general not complete, as now $S^{\downarrow}$ is only an under-approximation of any supervisor that is consistent with $O$~\cite{LTZS20}.
%\emph{Proof:} Similar to the proof of \emph{Theorem IV.3} and \emph{Theorem IV.4}, it can be checked that the sensor-actuator attacker $A$ generated in \textbf{Procedure 2}, if non-empty, is a sound solution for \textbf{Problem 1}. \hfill $\blacksquare$

\begin{figure}[htbp]
\begin{center}
\includegraphics[height=8.5cm]{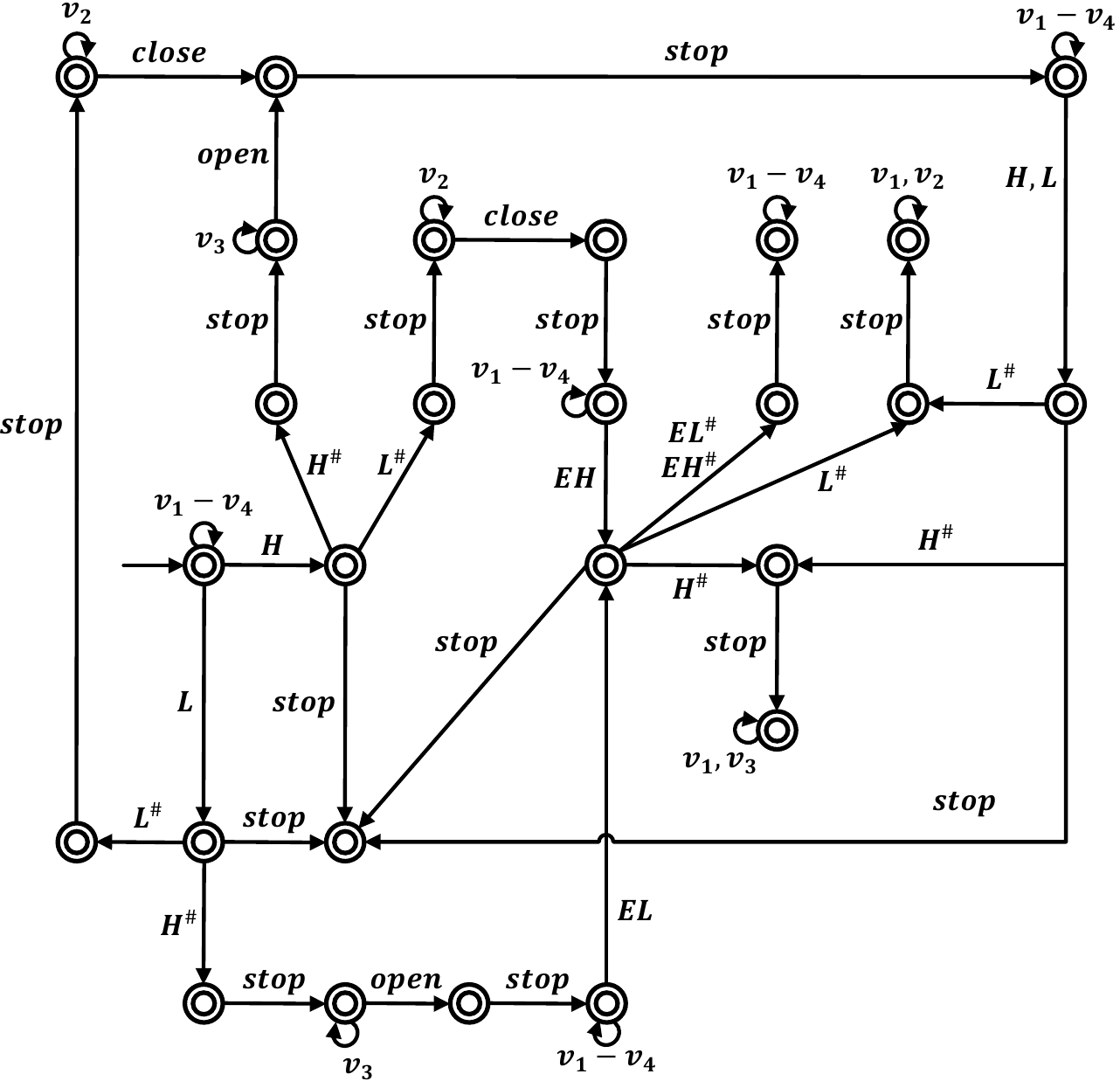}   
\caption{The synthesized sensor-actuator attack $A$}
\label{fig:Example_A}
\end{center}        
\end{figure}

\textbf{Example IV.4} We shall continue with the water tank example. Based on \textbf{Procedure 2}, %plant $G$, command execution automaton $CE^{A}$ under actuator attack, attack constraints $AC$, constructed $OCNS^{A}$ and $S^{\downarrow,A}$ shown in Fig. \ref{fig:Plant G}, \ref{fig:Example_command execution}, \ref{fig:Example_Sensor attack constraints AC}, \ref{fig:Example_OCNS^{A}}, \ref{fig:Example_S_downarrow_A}, respectively, 
we can use \textbf{SuSyNA} \cite{Susyna} to synthesize the supremal covert damage-reachable sensor-actuator attacker $A$, which is illustrated in Fig. \ref{fig:Example_A}.

Intuitively, the attack strategy of the synthesized sensor-actuator attacker $A$ is explained as follows. Upon the observation of $H$ ($L$, respectively), it would immediately replace it with the fake sensing information $L^{\#}$ ($H^{\#}$, respectively). Once the supervisor receives this fake information, it will issue the inappropriate control command $v_{2}$ ($v_{3}$, respectively). When the water tank system receives such a control command, it will execute the event $close$ ($open$, respectively), resulting in that the water level becomes $EH$ ($EL$, respectively), that is, the damage-infliction goal is achieved. In addition, such an attack strategy would always allow the attacker to remain covert against any safe supervisor that is consistent with the observations $O$ shown in Fig. \ref{fig:Observations M_o}.
%{\color{red} Indeed, for this example, the observation set $O=\{\epsilon\}$ is also fine for this approach, while the approach in [24] would not work. We can briefly discuss on the new example at the end of the paper to illustrate the advantage of the new approach, compared with the old one (as currently the example used are more or less the same) and I am not sure if the end result attacker is the same expressive or not.}{\color{blue}I think maybe there is no solution based on the Fig. \ref{fig:Example_S_To_BT(S)A} because once the supervisor observes $close$ or $open$, the covertness is broken while the damage state has not been reached, thus, the attacker cannot achieve the damage goal.}
%{\color{red} one may also be interested in the limit of observation-assisted approach. As our approach is not limited to finite observations.... A fundamental problem to ask is how many observations do we need to observe? do we need to observe continously if it keeps failing the synthesis? We show that in the limit we can decide this problem...only need to resolve the deadlock state in observation automaton...}
%{\color{red} may add a dump state to deal with infinite observations..to $M_o$}

%%%%%%%%%%%%%%%%%%%%%%%%%%%%%%%%%%%%%%%%%%%%%%%%%%%%%%%%%%%%%%%%%%%%%%%%%%%%%%%%%%%%

\section{Conclusions}
\label{sec:Conclusions}
This work investigates the problem of synthesizing maximally permissive covert sensor-actuator attackers to ensure damage reachability against unknown supervisors, where only a finite set of collected observations instead of the supervisor model is needed. We have shown the decidability of the observation-assisted covert  damage-reachable attacker synthesis problem. Our solution methodology is to reduce the original problem into the Ramadge-Wonham supervisory control problem, which allows several existing synthesis tools \cite{Susyna}-\cite{Malik07} to be used for the synthesis of covert  damage-reachable attackers against unknown supervisors. In the future works, we shall relax the normality assumption and study the decidability problem, and explore more powerful synthesis approaches to achieve the damage-nonblocking goal against unknown supervisors.

%%%%%%%%%%%%%%%%%%%%%%%%%%%%%%%%%%%%%%%%%%%%%%%%%%%%%%%%%%%%%%%%%%%%%%%%%%%%%%%%%%%%

\begin{IEEEbiography}[
{
\includegraphics[width=1.0in,height=1.40in,clip,keepaspectratio]{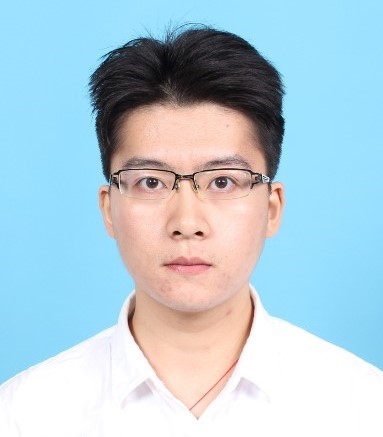}
}
]
{Ruochen Tai}
received the B.E. degree in electrical engineering from the Nanjing University of Science and Technology in 2016, and M.S. degree in automaton from the Shanghai Jiao Tong University in 2019. He is currently pursuing the Ph.D. degree with Nanyang Technological University, Singapore. His current research interests include security issue of cyber-physical systems, multi-robot systems, safe autonomy in cyber-physical-human systems, formal methods, and discrete-event systems.
%systems.
\end{IEEEbiography}
\begin{IEEEbiography}[
{
\includegraphics[width=1.0in,height=1.40in,clip,keepaspectratio]{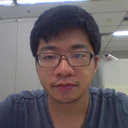}
}
]
{Liyong Lin}
received the B.E. degree and Ph.D. degree in electrical engineering in 2011 and 2016, respectively, both from Nanyang Technological University, where he has also worked as a project officer. From June 2016 to October 2017, he was a postdoctoral fellow at the University of Toronto. Since December 2017, he has been working as a research fellow at the Nanyang Technological University. His main research interests include supervisory control theory and formal methods. He previously was an intern in the Data Storage Institute, Singapore, where he worked on single and dual-stage servomechanism of hard disk drives.
%systems.
\end{IEEEbiography}
\begin{IEEEbiography}[
{
\includegraphics[width=1.0in,height=1.20in,clip,keepaspectratio]{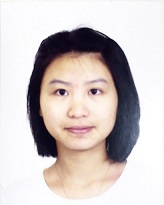}
}
]
{Yuting Zhu}
received the B.S. degree from Southeast University, Jiangsu, China, in 2016. She is currently pursuing the Ph.D. degree with Nanyang Technological University, Singapore. Her research interests include networked control and cyber security of discrete event systems.
%systems.
\end{IEEEbiography}
\begin{IEEEbiography}[
{
\includegraphics[width=1in,height=1.3in,clip,keepaspectratio]{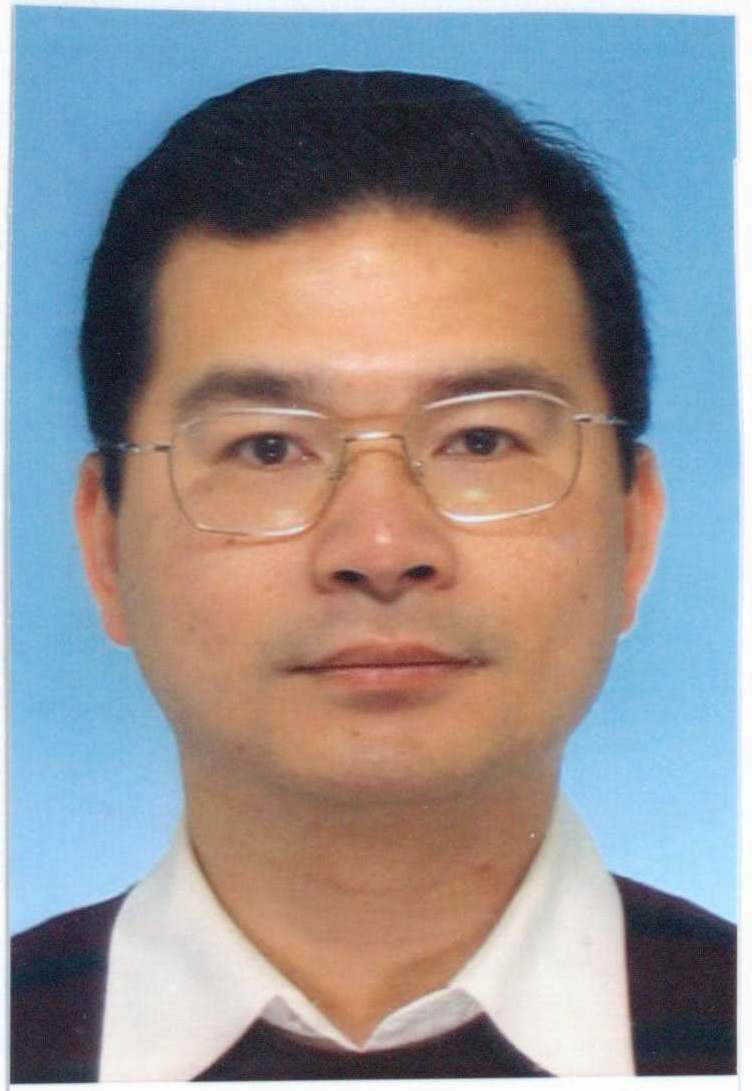}
}
]
{Rong Su} received the Bachelor of Engineering degree from University of Science and Technology of China in 1997, and the Master of Applied Science degree and PhD degree from University of Toronto, in 2000 and 2004, respectively. He was affiliated with University of Waterloo and Technical University of Eindhoven before he joined Nanyang Technological University in 2010. Currently, he is an associate professor in the School of Electrical and Electronic Engineering. Dr. Su's research interests include multi-agent systems, cybersecurity of discrete-event systems, supervisory control, model-based fault diagnosis, control and optimization in complex networked systems with applications in flexible manufacturing, intelligent transportation, human-robot interface, power management and green buildings. In the aforementioned areas he has more than 220 journal and conference publications, and 5 granted USA/Singapore patents. Dr. Su is a senior member of IEEE, and an associate editor for Automatica, Journal of Discrete Event Dynamic Systems: Theory and Applications, and Journal of Control and Decision. He was the chair of the Technical Committee on Smart Cities in the IEEE Control Systems Society in 2016-2019, and is currently the chair of IEEE Control Systems Chapter, Singapore.

\end{IEEEbiography}

\end{document}